\documentclass[a4paper,12pt]{article}
\usepackage{graphicx,amssymb,bm,latexsym}
\makeatletter
\@addtoreset{equation}{section}

\makeatother
\newcommand{\bea}{\begin{eqnarray}}
\newcommand{\ena}{\end{eqnarray}}
\newcommand{\vs}[1]{\vspace{#1 mm}}
\newcommand{\hs}[1]{\hspace{#1 mm}}
\renewcommand{\a}{\alpha}
\renewcommand{\b}{\beta}
\renewcommand{\c}{\gamma}

\newcommand{\e}{\epsilon}
\newcommand{\s}{\sigma}

\newcommand{\dsl}{\pa \kern-0.5em /}

\newcommand{\pa}{\partial}

\newcommand{\nn}{\nonumber\\}
\newcommand{\p}[1]{(\ref{#1})}

\newcommand{\dddot}[1]{\stackrel{...}{#1}}
\begin{document}

\begin{titlepage}

\begin{flushright}
OU-HET 497 \\
WU-AP/200/04\\
hep-th/0411093
\end{flushright}

\vs{10}
\begin{center}
{\Large\bf Inflation from Superstring/M Theory Compactification with
Higher Order Corrections I}
\vs{10}

{\large
Kei-ichi Maeda$^{a,b,c,}$\footnote{e-mail address:
maeda@gravity.phys.waseda.ac.jp} and Nobuyoshi Ohta$^{d,}$\footnote{e-mail
address: ohta@phys.sci.osaka-u.ac.jp}}\\
\vs{10}
$^a${\em Department of Physics, Waseda University,
Shinjuku, Tokyo 169-8555, Japan}\\
$^b$ {\em Advanced Research Institute for Science and Engineering,
Waseda University, Shinjuku, Tokyo 169-8555, Japan~} \\
$^c$ {\em Waseda Institute for Astrophysics, Waseda University,
Shinjuku, Tokyo 169-8555, Japan~}  \\
$^d${\em Department of Physics, Osaka University,
Toyonaka, Osaka 560-0043, Japan}
\end{center}
\vs{10}

\centerline{{\bf{Abstract}}}
\vs{5}

We study time-dependent solutions in M and superstring theories with
higher order corrections. We first present general field equations
for theories of Lovelock type with stringy corrections in arbitrary
dimensions. We then exhaust all exact and asymptotic solutions of
exponential and power-law expansions in the theory with Gauss-Bonnet
terms relevant to heterotic strings and in the theories with quartic
corrections corresponding to the M-theory and type II superstrings.
We discuss interesting inflationary solutions
that can generate enough e-foldings in the early universe.

\end{titlepage}
\newpage
\renewcommand{\thefootnote}{\arabic{footnote}}
\setcounter{footnote}{0}
\setcounter{page}{2}

\section{Introduction}

There are two major questions that confront the current cosmology.
One is the horizon problem which asks why the early universe
is highly homogeneous beyond causally disconnected regions.
The other is the flatness problem: why does the present universe appear
so extremely flat? It is widely believed that these can be resolved
once we accept that our universe underwent an inflationary evolution
in the early epoch. The recent cosmological observation confirmed
the existence of the early inflationary cosmological epoch as well as
the accelerated expansion of the present universe~\cite{WMAP}.

Though it is not difficult to construct cosmological models with these
features if one introduces scalar fields with suitable potentials,
it is desirable to derive such a model from the fundamental
theories of particle physics that incorporate gravity without making
special assumptions in the theories. The most promising candidates for
such theories are the ten-dimensional superstrings or eleven-dimensional
M-theory, which are hoped to give models of accelerated
expansion of the universe upon compactification to four dimensions.

However it was believed that there was a ``no-go'' theorem which forbids
such solutions if the six- or seven-dimensional internal space is
a time-independent nonsingular compact manifold without boundary~\cite{NG}.
A progress has recently been made by the discovery that this no-go theorem
can be evaded if the size of the internal space depends on time.
In fact, it has been shown that a model with certain period of accelerated
expansion can be obtained from the higher-dimensional vacuum Einstein
equation if one takes the internal space hyperbolic and its size depending
on time~\cite{TW}. It has been shown~\cite{NO} that this class of models
is obtained from what are known as S-branes~\cite{Sbrane1,Sbrane2}
in the limit of vanishing flux of three-form fields (see also~\cite{Sbrane3}).
It is found that this wider class of solutions give accelerating
universes for internal spaces with zero and positive curvature as well
if the flux is introduced. Further discussion of this class of models has
been given in Refs.~\cite{W}-\cite{exp}.

It turns out, however, that the models thus obtained do not give enough
e-foldings necessary to explain the cosmological problems mentioned
above~\cite{NO,W}. The reason for this can be understood from the
viewpoint of the effective four-dimensional theory,
where one gets gravitational theory coupled to scalar fields which
characterize the sizes of the internal spaces. Typically one finds
exponential potential, and the slope for the scalar fields in this
potential is too steep to produce large enough expansion~\cite{cosm2}.
Some efforts to overcome this problem were made in the present framework
in Ref.~\cite{cosm3}.

The scale when the acceleration occurs in this type
of models is basically governed by the Planck scale in the higher ten
or eleven dimensions. With phenomena at such high energy, it is expected
that we cannot ignore higher order corrections such as higher derivative
terms in the theories at least in the early universe.
It is known that there are terms of higher orders in the curvature to
the lowest effective supergravity action coming from superstrings or
M-theory~\cite{Be,hetero,TBB}.
With such corrections, they will significantly affect the inflation
at the early stage of the evolution of our universe.

The cosmological models in higher dimensions were studied intensively
in the 80's by many authors~\cite{KK_cosmology}-\cite{HAL}.
It was shown that inflation is indeed possible with higher-order
curvature corrections~\cite{old1,ISHI}. (The no-go theorem does
not apply to theories with higher derivatives.)
In particular the model with the Gauss-Bonnet (GB) terms is interesting
because they are the special combination without ghost~\cite{Zw}
and they exist as higher order corrections in the heterotic string
theories~\cite{Be}.
It was shown that there are two exponentially expanding solutions,
which may be called generalized de Sitter solutions since the size of the
internal space also depends on time (otherwise there is no solution of this
type)~\cite{ISHI}.
In both solutions, the external space inflates, while the internal space
shrinks exponentially. (There are also two time-reversed solutions, i.e. the
external space shrinks exponentially but the internal space inflates.)
One solution is stable and the other is unstable. Since the present
universe is not in the phase of de Sitter expansion with this energy scale,
we cannot use the stable solution for a realistic universe. If we adopt
the unstable solution, on the other hand, we may not find sufficient
inflation unless we fine-tune the initial values. The higher-order
curvature terms called Lovelock gravity~\cite{DF} and other types~\cite{ABF}
were also considered in higher-dimensional cosmology.

A good news is that large e-folding number was obtained in these models.
However, most of the work considered powers of scalar curvature or
Lovelock gravity, which are not the types of corrections known to
arise in superstring theories or M-theory. It is thus important to
examine if the above result of small e-folding is modified with higher-order
corrections expected in these fundamental theories.
In our previous paper~\cite{MO}, we have presented brief report of our
results on this problem for M-theory. Here we give the details of our
results in M-theory as well as superstrings. We focus on the solutions to
the vacuum Einstein equations with higher order corrections
since the basic feature can be obtained in this simple setting.
In this paper, we exhaust exact solutions as well as past and future
asymptotic solutions and discuss inflationary solutions among them.
The past and future asymptotic solutions are useful in describing
the inflation at the early universe and the present accelerating cosmology,
respectively. In a forthcoming paper~\cite{AMO}, we shall discuss more
detailed properties of these solutions including stability and
possible scenario for the history of our universe.

In the next section,
we present our actions and field equations to be solved. We write down
these for $D=1+p+q$ dimensions with $p$ external and $q$ internal space
dimensions. Though we are mainly interested in $p=3$ in this paper,
there may be interesting applications if we keep the dimension $p$ arbitrary.
Also we give the equations for maximally symmetric spaces with nonvanishing
curvatures. Their explicit forms are given in Appendices A -- C.
The Lovelock part of the field equations generalizes those
in Ref.~\cite{DF} and should be useful for examining various other nontrivial
solutions. We also discuss the relation of the solutions to those in the
Einstein frame in $(p+1)$ dimensions.

In Sec.~3, we examine solutions to the vacuum Einstein equations with
GB corrections, corresponding to the heterotic strings~\cite{Be,hetero}.
We exhaust possible generalized de Sitter and power-law solutions, and
find inflationary models for several types of internal
spaces with positive, zero and negative curvatures.
We find that exponential type solutions are possible for flat external
and internal spaces, corresponding to those solutions obtained in
the 80's by Ishihara~\cite{ISHI}.

In type II superstrings or M-theory, it is known that the coefficient of
the GB terms is zero and the first higher corrections start
with $R^4$ terms~\cite{TBB}. We study this case and find
interesting solutions of exponential and power-law expansions in Sec.~4.

Finally in Sec.~5, we summarize our solutions and discuss inflationary
solutions. We find that some solutions do not give inflations in
the Einstein frame in four dimensions even though they appear to give
inflations in the original frame, and that there are peculiar cases
in which inflation appears to be realized in the Einstein frame though
the external space is shrinking in the original frame.
We discuss which solutions are suitable for interesting cosmologies.

\section{Vacuum Einstein equations with higher order terms}

We consider the low-energy effective action for superstrings ($D=10$) or
M-theory ($D=11$) with higher order corrections in $D$ dimensions:
\bea
S &=& \sum_{n=1}^4 S_n+S_S,
\label{totaction}
\ena
with
\bea
\label{eh}
S_1 &=&S_{\rm EH} ~\equiv~\frac{\a_1}{2\kappa_{D}^2} \int d^{D} x
\sqrt{-g} R,\\
\label{gb}
S_2 &=&S_{\rm GB} ~\equiv~ \frac{\a_2}{2\kappa_{D}^2} \int d^{D} x
\sqrt{-g}\;  \left[R_{\mu\nu\rho\sigma}^2 - 4 R_{\mu\nu}^2 +R^2\right],\\
S_3 &=&
\frac{\a_3}{2\kappa_{D}^2} \int d^{D} x
\sqrt{-g}\;  \tilde{E}_{6}\,, \\
S_4 \! &=&  \frac{\a_4}{2\kappa_{D}^2}\int d^{D} x
\sqrt{-g}  ~\tilde{E}_8  \,,
 \\
S_S \! &=&  \frac{\c}{2\kappa_{D}^2}\int d^{D} x
\sqrt{-g} ~\tilde{J}_0  \,,
\label{4th1}
\ena
where
\bea
\tilde{E}_{2n} &=&\!\!\! -{1\over 2^n  (D-2n)!}
\e^{\a_1 \cdots \a_{D-2n} \mu_1 \nu_1 \ldots \mu_n \nu_n}
\e_{\a_1 \cdots \a_{D-2n} \rho_1 \sigma_1 \ldots \rho_n \sigma_n}
R^{\rho_1\sigma_1}{}_{\mu_1 \nu_1}
\cdots R^{\rho_n \sigma_n}{}_{\mu_n \nu_n}
\,, \\
\tilde{J}_0&=& R^{\lambda\mu\nu\kappa}R_{\a\mu\nu\b}
R_{\lambda}{}^{\rho\sigma\a} R^\b{}_{\rho\sigma\kappa}
+\frac12 R^{\lambda\kappa\mu\nu}R_{\a\b\mu\nu}R_{\lambda}{}^{\rho\sigma\a}
R^\b{}_{\rho\sigma\kappa}.
\ena
Here we have dropped contributions from forms and dilatons (if they exist),
$\kappa_{D}^2$ is a $D$-dimensional gravitational constant,
and we leave the coefficients $\a_1,\ldots, \a_4$ and $\c$ free.
The coefficient $\a_1$ of the Einstein-Hilbert (EH) term is $1$ by
definition, and though $\a_3$ is zero for all superstrings and M-theory,
we have included it since it will be useful for examining other cases.
For the heterotic strings, the leading correction is given by the
GB terms with the coefficient~\cite{Be,hetero}:
\bea
\a_2=\frac{1}{8}\alpha',
\label{hetero}
\ena
multiplied by an exponential factor of the dilaton, where $\alpha'$ is
the Regge slope parameter. So it is not immediately obvious if our
solutions are valid within the full string theories. Nevertheless they
would give solutions for constant dilaton, and our results for these
cases should be understood with this restriction.
For the M-theory in 11 dimensions, the coefficient for the
GB terms $\a_2$ vanishes, so we should consider forth order
terms with the coefficients~\cite{TBB}:
\bea
\a_2 = \a_3 = 0, \quad
\a_4 = - {\kappa_{11}^2 ~T_2\over 3^2\times 2^{9} \times (2\pi)^4} ,\qquad
\gamma = - {\kappa_{11}^2 ~ T_2\over 3 \times 2^{4}\times (2\pi)^4} \,,
\label{m}
\ena
where $T_2=({2\pi^2 /\kappa_{11}^2})^{1/3}$ is the membrane tension.
Type II superstring has the same couplings as M-theory in 10 dimensions,
so we can discuss this case if we keep the dilaton field constant
and ignore the contributions from other fields.
We should also note that contributions of the Ricci tensor $R_{\mu\nu}$
and scalar curvature $R$ are not included in the fourth-order
corrections~(\ref{4th1}) because these terms are not uniquely fixed.

\subsection{Basic equations for cosmology}
\label{sec2.1}

Since we are interested in cosmological solutions, we take the metric of our
$D$-dimensional space as
\bea
ds_D^2 &=& -N^2(t)dt^2 + a^2(t) ds_{p}^2
+ b^2(t) ds_{q}^2,
\label{met1}
\ena
with
\bea
N(t) = e^{u_0(t)} , ~ a(t) = e^{u_1(t)} , ~b(t) = e^{u_2(t)},
\ena
where $D=1+p+q$.
The external $p$- and internal $q$-dimensional
spaces ($ds_{p}^2$ and $ds_{q}^2$) are chosen to be maximally
symmetric.
The curvature constants of those spaces are defined by
$\sigma_p$ and $\sigma_q$. The sign of $\sigma_p$ ($\sigma_q$)
determines the type of maximally symmetric spaces, i.e.  $\sigma_p$
(or $\sigma_q$) = $-1$, 0 and 1 denote a hyperbolic space, a flat
Euclidean space, and a sphere, respectively. The volumes of the hyperbolic
and flat spaces are made finite by dividing by discrete isometry.

{}From the variation of the total action~\p{totaction} with respect to
$u_0, u_1$ and $u_2$, we find three basic field equations:
\bea
\label{eq1}
&&
F\equiv \sum_{n=1}^4 F_n+F_S=0\,,
\\
\label{eq2}
&&
F^{(p)}\equiv \sum_{n=1}^4
f_n^{(p)}+X \sum_{n=1}^4
g_n^{(p)}+Y \sum_{n=1}^4
h_n^{(p)} +F_S^{(p)}=0\,,
\\
\label{eq3}
&&
F^{(q)}\equiv \sum_{n=1}^4
f_n^{(q)}+Y \sum_{n=1}^4
g_n^{(q)}+X\sum_{n=1}^4
h_n^{(q)} +F_S^{(q)}=0\,,
\ena
where $X=\ddot{u}_1-\dot{u}_0\dot{u}_1+\dot{u}_1^2$,
$Y=\ddot{u}_2-\dot{u}_0\dot{u}_2+\dot{u}_2^2$, and
\bea
&&F_n=F_n(u_0,\dot{u}_1,\dot{u}_2, A_p,A_q),
\nn &&
F_S=(u_0,u_1,u_2,\dot{u}_0,\dot{u}_1,\dot{u}_2,\ddot{u}_1,
\ddot{u}_2,\dddot{u}_1,\dddot{u}_2,X,Y,\dot{X},\dot{Y}),
\nn &&
f_n^{(p)}=f_n^{(p)}(u_0,\dot{u}_1,\dot{u}_2, A_p,A_q),
\nn &&
g_n^{(p)}=g_n^{(p)}(u_0,\dot{u}_1,\dot{u}_2, A_p,A_q),
\nn &&
 h_n^{(p)}= h_n^{(p)}(u_0,\dot{u}_1,\dot{u}_2, A_p,A_q),
\nn &&
 F_S^{(p)}= F_S^{(p)}(u_0,u_1,u_2,\dot{u}_0,\dot{u}_1,
\dot{u}_2,\ddot{u}_1,\ddot{u}_2,\dddot{u}_1,\dddot{u}_2,
X,Y,\dot{X},\dot{Y},\ddot{X},\ddot{Y}),
\nn &&
 f_n^{(q)}= f_n^{(q)}(u_0,\dot{u}_1,\dot{u}_2, A_p,A_q),
\nn &&
g_n^{(q)}=g_n^{(q)}(u_0,\dot{u}_1,\dot{u}_2, A_p,A_q),
\nn &&
 h_n^{(q)}= h_n^{(q)}(u_0,\dot{u}_1,\dot{u}_2, A_p,A_q),
\nn &&
F_S^{(q)}=F_S^{(q)}(u_0,u_1,u_2,\dot{u}_0,\dot{u}_1,
\dot{u}_2,\ddot{u}_1,\ddot{u}_2,\dddot{u}_1,\dddot{u}_2,
X,Y,\dot{X},\dot{Y},\ddot{X},\ddot{Y}),
\ena
are explicitly given in Appendix A.
Here $A_p$ and $A_q$ are defined by
\bea
&&
A_p=\dot{u}_1^2+\sigma_p \exp [2(u_0-u_1)],
\nn
&&
A_q=\dot{u}_2^2+\sigma_q \exp [2(u_0-u_2)].
\ena

Since $u_0$ is a gauge freedom of time coordinate, we have three equations
for two variables $u_1$ and $u_2$. It looks like an over-determinant system.
However, these three equations are not independent.
In fact, we can derive the following equation after bothersome calculation:
\bea
\dot{F}+(p\dot{u}_1+q\dot{u}_2-\dot{u}_0) F=
p\dot{u}_1  F^{(p)}+q\dot{u}_2  F^{(q)}\,.
\label{relation:FFpFq}
\ena
If $F=0$ and $F^{(p)}=0$ (or $F^{(q)}=0$), we obtain $\dot{u}_2 F^{(q)}=0$
(or $\dot{u}_1  F^{(p)}=0$), since $F=0$ is a constraint equation and
its time derivative also vanishes. The third equation $F^{(q)}=0$
(or $F^{(p)}=0$) is then automatically satisfied unless
$\dot{u}_2=0$ (or $\dot{u}_1=0$).
On the other hand, suppose we have only $F^{(p)}=0$ and $F^{(q)}=0$.
Then Eq.~(\ref{relation:FFpFq}) gives
\bea
F=C e^{u_0-(pu_1+qu_2)}\,,
\label{ini}
\ena
where $C$ is an integration constant. Upon imposing the initial condition
$F=0$ in \p{ini}, we get $C=0$ and hence $F=0$. This means that the constraint
equation is satisfied if other dynamical equations are solved {\it and}
it is initially satisfied. Consequently, it is in general not enough to solve
the dynamical equations $F^{(p)}=F^{(q)}=0$ only, but enough to solve the two
equations $F=0$ and $F^{(p)}=0$ (or $F^{(q)}=0$) instead of trying to
solve all three equations.

\subsection{Ansatz for solutions}

We now analyze our basic Eqs.~\p{eq1} -- \p{eq3} for several models and
look for inflationary solutions. Since we are interested in analytic
solutions, we study the following two cases:\\
{\bf (1) Generalized de Sitter solutions:}\\
Using a cosmic time, i.e. $u_0=0$, an exponential expansion of each scale
factor is given by $u_1=\mu t+\ln a_0$, and
$u_2=\nu t+\ln b_0$, where $\mu, \nu, a_0$ and $b_0$ are constants.\\
{\bf (2) Power-law solutions:}\\
To find a power-law solution, although we can discuss it with the above
cosmic time, we use a different time gauge, which is defined by
$u_0=t$. Using this time coordinate, a power-law solution is given by
$u_1=\mu t+\ln a_0$, and $u_2=\nu t+\ln b_0$, where $\mu$ and $\nu$ are
constants.

The choice of time coordinate in (2) is more convenient than the cosmic
time in (1) because we can discuss both solutions in the same set up.
Namely we can write
\bea
u_0=\epsilon t,\quad
u_1=\mu t+\ln a_0,\quad
\mbox{ and } \quad
u_2=\nu t+\ln b_0,
\label{ansatz}
\ena
where
$\epsilon = 0$ for case (1), while $\epsilon =1$ for case (2).
In the latter case, in terms of a new cosmic time $\tau=e^t$, we see
that the solution gives the power-law behavior:
\bea
a=e^{u_1}=a_0\tau^\mu, ~~~{\rm and}~~~b=e^{u_2}=b_0\tau^\nu \,.
\ena
Note that when the curvature constant $\sigma_p$ (or $\sigma_q$) vanishes,
$a_0$ and $b_0$ are arbitrary but we can set $a_0=1$ (or $b_0=1$) because
such a numerical constant can be absorbed by rescaling of the spatial
coordinates.

\subsection{Description in the Einstein frame}
\label{einframe}

After the internal space is compactified, we observe physical variables
in the $(1+p)$-dimensional Einstein frame, which is defined by
\bea
ds_D^2=e^{-2{q\over p-1}\phi}(-dt_E^2+a_E^2 ds_p^2)+e^{2\phi}ds_q^2\,,
\label{met2}
\ena
where $t_E$, $a_E$, and $\phi \,(=u_2=\ln b)$ are a cosmic time, a scale
factor, and a scalar field parametrizing the size of the internal space
in the Einstein frame, respectively.
Comparing Eqs. (\ref{met1}) and (\ref{met2}), we find the relations
\bea
&&e^{u_0}dt=\pm e^{-{q\over p-1}\phi}dt_E,
\label{time_trans}\\
&&e^{u_1}=e^{-{q\over p-1}\phi}a_E, \\
&&u_2=\phi\,.
\ena
The sign $\pm$ in Eq. (\ref{time_trans}) is chosen so that
two time coordinates proceed in the same (future) direction.
The solutions in the form~\p{ansatz}
can be rewritten in the Einstein frame as follows:
\begin{description}
\item[(1) $\epsilon=0$ and $\nu> 0$]
\bea
&&t_E=t_E^{(0)} \, e^{{q\over p-1}\nu t},
\label{time1}\\
&&a_E=a_E^{(0)}\left|{t_E\over t_E^{(0)}}\right|^\lambda,
\label{aE1} \\
&&\phi=\phi^{(0)}+{(p-1)\nu\over q}\ln\left|{t_E\over t_E^{(0)}}\right|\,,
\label{phi1}
\ena
where $t_E^{(0)} (>0) $, $a_E^{(0)}$ and $\phi^{(0)}$ are integration
constants, and
\bea
\lambda=1+{(p-1)\mu\over q\nu}\,.
\ena
If $\mu/\nu>0$, this solution gives a power-law inflation in the Einstein
frame.
$t=-\infty$ and $t=\infty$ correspond to $t_E=0$ and $t_E=\infty$,
respectively.

\item[(2) $\epsilon=0$ and $\nu< 0$]
\bea
t_E=t_E^{(0)} \, e^{{q\over p-1}\nu t},
\label{time2}
\ena
where $t_E^{(0)} (<0) $ is an integration constant, and
$a_E$ and $\phi$ are the same as Eqs. (\ref{aE1}) and (\ref{phi1}).
$t\in (-\infty,\infty)$ is transformed into $t_E\in (-\infty,0)$.
$t=-\infty$ and $t=\infty$ correspond to $t_E=-\infty$ and $t_E=0$,
respectively. In this case the inflationary solutions in the Einstein frame
are obtained for $\lambda<0$, i.e., when $t_E \rightarrow 0_-$, $a_E$
diverges as $|t_E|^{-|\lambda|}$. This is called a super inflation in
Kaluza-Klein cosmology~\cite{KK_cosmology,super_inflation}. Since the
asymptotic behaviour as $t_E\rightarrow 0$ does not explain the present
universe, we have to avoid a singularity at $t_E=0$.
Then, we have to clarify a mechanism to avoid the singularity at
$t_E=0$. The same problem was found in the Kaluza-Klein inflation in
80's~\cite{KK_cosmology}. In a pre-big bang scenario, we also find
a similar inflation in the string frame~\cite{pre_bigbang}.

~Note that even for $\mu>0$, this class of solutions in general
do not give inflationary solutions in the Einstein frame.

\item[(3) $\epsilon=0$ and $\nu= 0$]
\bea
&&t_E=e^{{q\over (p-1)}\phi^{(0)}}t, \\
&&a_E=a_E^{(0)}\exp\left[{\mu e^{-{q\over (p-1)}\phi^{(0)}} t_E}\right], \\
&&\phi=\phi^{(0)}\,,
\ena
where  $a_E^{(0)}$ and $\phi^{(0)}$ are constants.
Rescaling the time coordinate, we can set $\phi^{(0)}=0$, i.e. $t_E=t$
and $a_E\propto \exp (\mu t_E)$. This solution gives
an exponential expansion even in the Einstein frame for $\mu>0$.

\item[(4) $\epsilon=1$ and $\nu> -{p-1\over q}$]
\bea
&&t_E=t_E^{(0)}e^{\left(1+{q\over p-1}\nu\right) t},
\label{time3}\\
&&a_E=a_E^{(0)}\left|{t_E\over t_E^{(0)}}\right|^\lambda,
\label{aE2}\\
&&\phi=\phi^{(0)}+{(p-1)\nu\over (p-1)+q\nu}\ln\left|{t_E\over t_E^{(0)}}
\right|\,,
\label{phi2}
\ena
where $t_E^{(0)} (>0) $, $a_E^{(0)}$ and $\phi^{(0)}$ are integration
constants, and
\bea
\lambda={(p-1)\mu+q\nu\over (p-1)+q\nu}\,.
\ena
This solution gives a power-law inflation in the Einstein frame if $\mu>1$.
Remind that $\mu>1$ gives a power-law inflation in the original frame.
$t=-\infty$ and $t=\infty$ correspond to $t_E=0$ and  $t_E=\infty$,
respectively.

\item[(5) $\epsilon=1$ and $\nu< -{p-1\over q}$]
\bea
t_E=t_E^{(0)}e^{\left(1+{q\over p-1}\nu\right) t},
\ena
where $t_E^{(0)} (<0) $ is an integration constant, and
$a_E$ and $\phi$ are the same as Eqs.~(\ref{aE2}) and (\ref{phi2}).
$t\in (-\infty,\infty)$ is transformed into $t_E\in (-\infty,0)$.
$t=-\infty$ and $t=\infty$ correspond to $t_E=-\infty$ and $t_E=0$,
respectively. Here the inflationary solutions in the Einstein frame
are obtained for $\lambda<0$ (a super inflation).

\item[(6)  $\epsilon=1$ and $\nu= -{p-1\over q}$]
\bea
&&t_E=e^{{q\over (p-1)}\phi^{(0)}}t, \\
&&a_E=a_E^{(0)}\exp\left[{(\mu-1)  e^{-{q\over (p-1)}\phi^{(0)}} t_E}
\right],\\ &&\phi=\phi^{(0)}-{(p-1)\over q}e^{-{q\over (p-1)}\phi^{(0)}}
t_E\,,
\ena
where $\phi^{(0)}$ can be set zero by rescaling time coordinate, i.e.
$t_E=t$. This solution gives an exponential expansion in the Einstein frame
if $\mu>1$. Rescaling the time coordinate, we find
$t_E=t$, $a_E\propto \exp [(\mu-1) t_E]$, and  $\phi=-[(p-1)/q] t_E$.
\end{description}

These will be useful in discussing the results in the Einstein frame.

\section{Solutions in heterotic strings}

The higher order corrections for heterotic strings start with
the GB terms. So in this section, we first study various solutions
of the field equations only with EH and GB terms, which are given by
\bea
&&F_1+F_2=0,
\label{GB:basic1}\\
&&
F^{(p)}_1+F_2^{(p)}=0,
\label{GB:basic2}
\\
&&
F^{(q)}_1+F_2^{(q)}=0
\label{GB:basic3}
\,,
\ena
where
\bea
&&F_1=F_1(t, \epsilon,\mu,\nu,A_p,A_q), \nn
&&F_2=F_2(t, \epsilon,\mu,\nu,A_p,A_q), \nn
&&F_1^{(p)}=f_1^{(p)}(t, \epsilon,\mu,\nu,A_p,A_q)+X g_1^{(p)}(t,
\epsilon,\mu,\nu,A_p,A_q)+Y h_1^{(p)}(t, \epsilon,\mu,\nu,A_p,A_q), \nn
&&F_2^{(p)}=f_2^{(p)}(t, \epsilon,\mu,\nu,A_p,A_q)+X g_2^{(p)}(t,
\epsilon,\mu,\nu,A_p,A_q)+Y h_2^{(p)}(t, \epsilon,\mu,\nu,A_p,A_q), \nn
&&F_1^{(q)}=f_1^{(q)}(t, \epsilon,\mu,\nu,A_p,A_q)+Y g_1^{(q)}(t,
\epsilon,\mu,\nu,A_p,A_q)+X h_1^{(q)}(t, \epsilon,\mu,\nu,A_p,A_q), \nn
&&F_2^{(q)}=f_2^{(q)}(t, \epsilon,\mu,\nu,A_p,A_q)+Y g_2^{(q)}(t,
\epsilon,\mu,\nu,A_p,A_q)+X h_2^{(q)}(t, \epsilon,\mu,\nu,A_p,A_q)\,,\;\;\;\;
\ena
whose explicit expressions are given in Appendix B.
Here we have three equations for two unknown parameters $\mu$ and $\nu$.
However, two of them are independent because we have one constraint equation
(\ref{relation:FFpFq}).

{}From Eq.~(\ref{A_pA_q}), we expect there may exist no exact solution
except for the case of $\sigma_p=\sigma_q=0$.
However, even for the case of  $\sigma_p\neq 0$ or $\sigma_q\neq0$,
we may have some asymptotic analytic solutions either in the future
direction ($t\to \infty$) or in the past direction ($t\to -\infty$),
which describe cosmologies in these time regions.
We classify solutions to Eqs.~(\ref{GB:basic1}), (\ref{GB:basic2}), and
(\ref{GB:basic3}) by the signatures of $\sigma_p$ and $\sigma_q$.

\subsection{$\sigma_p=\sigma_q=0$}

In this case, $A_p=\mu^2$ and $A_q=\nu^2$ are constants.
We have two classes of solutions:\\
(1) exact solutions for $\epsilon=0$ (generalized de Sitter solutions),\\
(2) asymptotic solutions for $\epsilon=1$ (power-law solutions),\\
which are summarized below.

\subsubsection{Generalized de Sitter solutions ($\epsilon=0$)}

We have three basic equations one of which is a constraint equation.
In this case, however, as discussed in Appendix C, if the solution is not
the Minkowski space ($\mu=\nu=0$), we can find two independent algebraic
equations without any constraint equation:
\bea
&&F_1(\mu,\nu)+F_2(\mu,\nu)=0\,,
\label{GB:gdS_basic1}\\
&&
H_1(\mu,\nu)+H_2(\mu,\nu)=0\,,
\label{GB:gdS_basic2}
\ena
which are given in Appendix C, i.e.
\bea
&&\alpha_1 \left[p_1\mu^2+q_1\nu^2+2pq \mu\nu\right]
\nn&&
\hs{10} +\; \alpha_2 \left[ p_3\mu^4+6p_1q_1\mu^2\nu^2+q_3\nu^4+4\mu\nu\left(
p_2q\mu^2+pq_2\nu^2\right)\right]=0,
\label{H_eq}
\\
&&(\mu-\nu) \left\{\alpha_1 +2\alpha_2\left[
(p-1)_2\mu^2+2(p-1)(q-1)\mu\nu+(q-1)_2\nu^2 \right] \right\}=0\,.
\label{dynamical_eq1}
\ena
Now we have two branches of solutions: one is $\mu=\nu$, and the other is
\bea
\alpha_1
+2\alpha_2\left[
(p-1)_2\mu^2+2(p-1)(q-1)\mu\nu+(q-1)_2\nu^2 \right]
=0\,.
\label{2nd_order}
\ena

(1) Inserting $\mu=\nu$ into Eq.~(\ref{H_eq}), we find
either Minkowski space $\mu=\nu=0$, or another solution, i.e.
\bea
\mu=\nu=\pm \sqrt{-{\alpha_1(p_1+q_1+2pq)\over \alpha_2[p_3+q_3+6p_1q_1
+4(p_2q+pq_2)]}}\,,
\ena
if $\alpha_2<0$. For $\alpha_2>0$, there is no real solution.

(2) When we assume Eq.~(\ref{2nd_order}), eliminating $\alpha_2$ from
Eq.~(\ref{H_eq}), we obtain the fourth order equation:
\bea
&&(p+1)_2\mu^4+4p_1(p-1)(q-1)\mu^3\nu+2(p-1)(q-1)(3pq-2p-2q)\mu^2\nu^2
\nn
&&~~~~+4q_1(p-1)(q-1)\mu\nu^3
+(q+1)_2 \nu^4=0\,.
\label{4th}
\ena
If $\nu=0$, we have $\mu=0$, which gives Minkowski space.
Except for this trivial solution, Eq.~\p{4th} is reduced to the fourth
order equation for $h=\mu/\nu$:
\bea
&&(p+1)_2h^4+4p_1(p-1)(q-1)h^3+2(p-1)(q-1)(3pq-2p-2q)h^2
\nn
&&~~~~+4q_1(p-1)(q-1)h +(q+1)_2 =0\,.
\label{eq:h}
\ena
We then have four solutions for $h$. For a solution $h$ of this equation,
we get $\nu$ and then $\mu$ from Eq.~(\ref{2nd_order}), which is rewritten as
\bea
\alpha_1 +2\alpha_2\left[
(p-1)_2 h^2 +2(p-1)(q-1)h+(q-1)_2\right]\nu^2 =0\,.
\label{eq:nu}
\ena
We thus find
\bea
&&
\nu=\pm\sqrt{-{\alpha_1 \over 2\alpha_2\left[
(p-1)_2 h^2 +2(p-1)(q-1)h+(q-1)_2\right]}}\;,
\label{sol:nu}
\\&&
\mu=h\nu .
\label{sol:mu}
\ena
For $\nu$ to take a real value, we have a constraint
\bea
\alpha_2\left[
(p-1)_2 h^2 +2(p-1)(q-1)h+(q-1)_2)\right]<0.
\label{constraint_real}
\ena

For the heterotic strings with $\alpha_1=1$, $\alpha_2={\a'/8}$, $p=3$
and $q=6$, we have two real solutions for $h$ in Eq. (\ref{eq:h}).
Using these two solutions, we have the following four solutions
in the unit $\a'=1$:
\bea
(\mu,\nu) &=& (1.36601, -0.965665),~ (2.50608, -0.391608),
\label{sol1}
\ena
and the time-reversed ones $(-\mu,-\nu)$. In both solutions in (\ref{sol1}),
the external space inflates,
while the internal space shrinks exponentially. It was shown that one
solution is stable and the other is unstable~\cite{ISHI}. Since the present
universe is not in the phase of de Sitter expansion with this energy scale,
we cannot use the stable solution for a realistic universe. If we adopt
the unstable solution, on the other hand, we may not find sufficient
inflation unless we fine-tune the initial values. We shall also discuss
if these solutions give an inflation in the four-dimensional
Einstein frame in Sec.~5.

Though it is known that there is no GB terms for M-theory,
it may be instructive to find solutions for $p=3$ and $q=7$:
\bea
(\mu,\nu) = (1.45839, -0.838657),~ (2.53838, -0.331212),
\ena
and the time-reversed ones $(-\mu,-\nu)$.
Thus we find that the result does not change qualitatively.

\subsubsection{Power-law solutions ($\epsilon =1$)}
\label{gbe1}

Setting $\epsilon =1$ in Eqs. (\ref{gbeq1}) and (\ref{gbeq2}),
we find that the EH action is dominant as
$t\to \infty$, while the GB action becomes
dominant as $t\to -\infty$.
Here we present asymptotic power-law solutions for each case.

{\bf (1) Future asymptotic solutions ($t\to \infty$):}\\
Our basic equations reduce to
\bea
&& p_1\mu^2+q_1\nu^2+2pq \mu\nu =0,
\\
&&
q\nu(\nu-\mu-1)-(p-1)\mu=0, \\
&&
p\mu(\mu-\nu-1)-(q-1)\nu=0\,.
\ena
We can easily show that these three equations are equivalent to
the following two equations, if it is not Minkowski space ($\mu=\nu=0$):
\bea
p\mu^2+q\nu^2=1, ~~~
p\mu+q\nu=1\,,
\label{Kasner}
\ena
which is a special case of Kasner solutions. We have a solution
\bea
&&\mu={p\pm\sqrt{pq(p+q-1)}\over p(p+q)},
\nn
&&\nu={q\mp\sqrt{pq(p+q-1)}\over q(p+q)}\,.
\label{Kasner1}
\ena
For $p=3$ and $q=6$, we find $(\mu,\nu)=(5/9, -1/9), (-1/3, 1/3)$.
They are also future asymptotic solutions for type II superstrings.
Note that a general Kasner solution is given by
\bea
\sum_{i=1}^{p+q}\mu_i^2=1
, ~~~\sum_{i=1}^{p+q}\mu_i=1
\,,
\ena
where each scale factor is assumed as $e^{u_i}=\tau^{\mu_i}$ ($i=1, \cdots,
p+q$). Apparently $\mu<1$ in this class of solutions and they do not
give inflation according to the discussions in subsection~\ref{einframe}.

{\bf (2) Past asymptotic solutions ($t\to -\infty$):}\\
Our equations are
\bea
&&
p_3\mu^4+6p_1q_1\mu^2\nu^2+q_3\nu^4+4\mu\nu\left(
p_2q\mu^2+pq_2\nu^2\right)=0,
\label{eq:power_GB1}
\\
&&
q\nu\left[(q-1)_2\nu^3+(q-1)(2p-q)\mu\nu^2+(p-1)(p-2q)\mu^2\nu-(p-1)_2\mu^3
\right] \nn
&&
-\left[
q_2\nu^3+3(p-1)q_1\mu\nu^2+3(p-1)_2q\mu^2\nu+(p-1)_3\mu^3\right] =0,
\label{eq:power_GB2}
\\
&&
p\mu\left[(p-1)_2\mu^3-(p-1)(p-2q)\mu^2\nu-(q-1)(2p-q)\mu\nu^2-(q-1)_2\nu^3
\right]\nn
&&
-\left[
p_2\mu^3+3p_1(q-1)\mu^2\nu+3p(q-1)_2\mu\nu^2+(q-1)_3\nu^3\right] =0.
\label{eq:power_GB3}
\ena
We can show that Eq. (\ref{eq:power_GB1}) is derived from
Eqs.~(\ref{eq:power_GB2}) and (\ref{eq:power_GB3}), and
these three equations are not independent.
We can use any two of them to find the solutions.

We obtain the following equation from the difference between
Eqs.~(\ref{eq:power_GB2}) and (\ref{eq:power_GB3}):
\bea
&& (p\mu+q\nu-3)\left[ (p-1)_2\mu^3-(p-1)(p-2q)\mu^2\nu \right. \nn
&& \hs{30} \left.-(q-1)(2p-q)\mu\nu^2-(q-1)_2\nu^3\right]=0.
\ena
Thus we have either
\bea
p\mu+q\nu-3=0,
\ena
or
\bea
(p-1)_2\mu^3-(p-1)(p-2q)\mu^2\nu-(q-1)(2p-q)\mu\nu^2-(q-1)_2\nu^3=0.
\label{eq:power_GB4}
\ena

\noindent
(1): $p\mu+q\nu-3=0$\\
Here $\nu=0$ gives $\mu=3/p$, which is incompatible with Eq.~\p{eq:power_GB1}.
Thus $\nu\neq 0$, and Eq.~(\ref{eq:power_GB1}) is rewritten by $h=\mu/\nu$:
\bea
p_3 h^4 +
4p_2qh^3+6p_1q_1 h^2 +4pq_2h+q_3=0\,.
\label{eq:h2}
\ena
Once we find the solution of this fourth order equation, $\mu$ and $\nu$ are
given as
\bea
\mu={3h\over ph+q}, ~~~\nu={3 \over ph+q}.
\ena

If $p=3$, Eq.~(\ref{eq:h2}) reduces to a third order equation.
We can formally find three solutions for $h$ as
\bea
h = -{q-1 \over 2}+\sqrt{q^2-1\over 3}\cos\left[{1\over 3}\tan^{-1}
\left({1\over q}\sqrt{q^2-4\over 3}\right)+{2\pi n\over 3}\right]; \quad
n=0,1,2.
\ena

For the heterotic strings with $p=3$ and $q=6$, we find three real solutions:
\bea
&&h=-5.86861,~~~(\mu,\nu)=(1.51698, -0.25849),\nn
&&h=-1.30495,~~~(\mu,\nu)=(-1.87748, 1.43874),\nn
&&h=-0.32645,~~~(\mu,\nu)=(-0.19506, 0.59753).
\label{gbpower}
\ena
The first solution gives an inflation and is interesting.

\noindent
(2) Eq.~(\ref{eq:power_GB4})\\
Using $h=\mu/\nu$, Eq.~(\ref{eq:power_GB4}) reduces to a third order equation:
\bea
&&(p-1)_2h^3-(p-1)(p-2q)h^2 -(q-1)(2p-q)h -(q-1)_2 \nn
&&=
(h-1)\left[(p-1)_2h^2+2(p-1)(q-1)h+(q-1)_2\right]=0\,.
\label{eq:power_GB5}
\ena
We then have either $h=1$, or
\bea
h={1\over (p-1)(p-2)}\left[-(p-1)(q-1)\pm\sqrt{(p-1)(q-1)(p+q-3)}\right]\,.
\ena
However those are not consistent with Eq. (\ref{eq:power_GB1}).
Hence we have no solution in this case.

\subsection{$\sigma_p=0, \sigma_q\neq 0$ (or $\sigma_p\neq 0,  \sigma_q=0$)}

\subsubsection{Generalized de Sitter solutions ($\epsilon=0$)}

Since $A_p=\mu^2$ and $A_q=\nu^2+\tilde{\sigma}_q e^{-2\nu t}$, it is
easy to see that
$\nu$ must vanish for the existence of exact solutions, where
$\tilde{\sigma}_q=\sigma_q/b_0^2$. (We also introduce
$\tilde{\sigma}_p=\sigma_p/a_0^2$ for further calculations). Setting
$\nu=0$, we have
\bea
F &=& F_1+F_2, \nn
&=& \alpha_1
[p_1\mu^2+q_1\tilde{\sigma}_q]+\alpha_2[p_3\mu^4+2p_1q_1\mu^2\tilde{\sigma}_q
+q_3\tilde{\sigma}_q^2]=0, \nn F^{(p)} &=&
f_1^{(p)}+f_2^{(p)}+(g_1^{(p)}+g_2^{(p)})X+(h_1^{(p)} +h_2^{(p)})Y \nn
&=&
\alpha_1[p_1\mu^2+q_1\tilde{\sigma}_q]+\alpha_2[p_3\mu^4+2p_1q_1\mu^2
\tilde{\sigma}_q+q_3\tilde{\sigma}_q^2]=0, \nn F^{(q)}&=&
f_1^{(q)}+f_2^{(q)}+(g_1^{(q)}+g_2^{(q)})X+(h_1^{(q)}+h_2^{(q)})Y
\nn
&=& \alpha_1 [(p+1)_0\mu^2+(q-1)_2\tilde{\sigma}_q] \nn
&& \hs{5} +\; \alpha_2[(p+1)_2\mu^4+2(p+1)_0(q-1)_2\mu^2\tilde{\sigma}_q
+(q-1)_4\tilde{\sigma}_q^2]=0.
\label{eq:nu0}
\ena
Although the first and second equations are identical, the third one is
different. Since we have two undetermined variables
$\mu$  and $\tilde{\sigma}_q$ for two independent equations,
we may have some solutions. However, we find that there is
no real solution at least for $p=3$ and $q=6$.

If $\nu\neq 0$, we have only asymptotic solutions. If $\nu>0 ~(<0)$,
\bea
&&
~A_q\to \nu^2~~~~~~~~{\rm as}~~
t\to +\infty ~(-\infty),
\label{asymptotic+}\\
&&
A_q\to \tilde \sigma_q e^{-2\nu t}~~~{\rm as}~~
t\to -\infty ~(+\infty)\,.
\label{asymptotic-}
\ena
Then, for the case (\ref{asymptotic+}), as $t\to +\infty ~(-\infty)$,
we recover the previous generalized de Sitter solutions~\p{sol1}.
For the heterotic strings, we find that
\bea
&&
(\mu,\nu)\sim (-1.36601, 0.965665),~ (-2.50608, 0.391608),~~{\rm as}~~
t\to +\infty , \nn
&&
(\mu,\nu)\sim (1.36601, -0.965665),~ (2.50608, -0.391608),~~{\rm as}~~
t\to -\infty \,.
\ena
On the other hand, for the case (\ref{asymptotic-}), as $
t\to -\infty ~(+\infty)$, we do not find any asymptotic solutions
within our ansatz for solutions. This does not mean that there is
no time-dependent solution to this system but simply implies that there is no
solution within our ansatz. We can study the evolution of the system by
a numerical analysis.

In the case of $\sigma_p\neq 0$ and $\sigma_q= 0$,
we can obtain our result by exchanging $p, \mu$ and $q, \nu$.
We have only asymptotic solutions. For the heterotic strings,
we find that
\bea
&&
(\mu,\nu)\sim (1.36601, -0.965665),~ (2.50608, -0.391608),~~{\rm as}~~
t\to +\infty , \nn
&&
(\mu,\nu)\sim (-1.36601, 0.965665),~ (-2.50608, 0.391608),~~{\rm as}~~
t\to -\infty \,.
\ena
There are solutions in which our space inflates and internal space
shrinks at late times, but no such solutions at early era.

\subsubsection{Power-law solutions ($\epsilon=1$)}

Next we turn to the power-law solutions.
Let us classify the solutions into three cases depending on $\nu$:\\
{\bf (1) $\nu>1$}

In this case, as $t\to \infty$, the EH term becomes dominant and
we obtain the asymptotic solution in the previous subsection~\ref{gbe1}.
However no solutions satisfy the condition $\nu>1$ (see Eq. (\ref{Kasner})).
Thus there is no asymptotic solution of our form.
As $t\to -\infty$, the GB curvature terms become dominant,
but we find no consistent solution since $A_q$ diverges without any
balancing term.\\
{\bf (2) $\nu<1$}

As $t\to \infty$ with EH dominance, we again find no consistent solution.
As $t\to -\infty$ with GB dominance, we obtain the asymptotic
solutions~\p{gbpower} in the previous section. For the heterotic strings,
imposing the condition of $\nu<1$,
we find two  solutions, which are
$(\mu,\nu)=(1.51698, -0.25849)$, and $(-0.19506, 0.59753)$.\\
{\bf (3) $\nu=1$}

This case is a little bit special because both $A_p=\mu^2$ and
$A_q=1+\tilde{\sigma}_q$ are constants.
Then the time dependence in the basic equations
(\ref{gbeq1}) and (\ref{gbeq2}) is only $e^{-t}$
from the EH action and $e^{-3t}$ from the GB action.
In the future asymptotic solutions, as $t\to \infty$, the
EH term becomes dominant, and we are left with
\bea
&&p_1\mu^2+q_1(1+\tilde{\sigma}_q)+2pq \mu=0, \\
&&
(p+q-1)\mu=0, \\
&&
2p\mu(\mu-2)-2(q-1)(1+\tilde{\sigma}_q)=0\,.
\ena
We then have an asymptotic solution
$\mu=0$, $\nu=1$ and $\sigma_q=-1$ ($b_0=1$).
This is just Minkowski spacetime with Milne-type time slicing.
We find that this solution is also consistent with the GB term
because $A_p=0,A_q=0$ and $\mu=0$.
Hence this Minkowski solution is an exact one to the whole system.

Though this appears a rather trivial solution in the frame we are discussing,
it gives power-law solutions in the Einstein frame in $(p+1)$ dimensions
and a nontrivial solution, as discussed in subsection~\ref{einframe}.
Unfortunately the scale factor behaves like $a_E \sim t_E^{q/(p-1+q)}$
and it is not an inflationary solution.

In the past asymptotic region $t\to -\infty$, the GB term
becomes dominant. We have
\bea
&& \hs{-10}
p_3\mu^4+4p_2q \mu^3+4p_1q_1\mu^2+ 2p_1q_1\mu^2(1+\tilde{\sigma}_q)
+4pq_2\mu(1+\tilde{\sigma}_q)+q_3(1+\tilde{\sigma}_q)^2=0,
\label{eq:power2_GB1}
\\
&& \hs{-10}
(p+q-3)\mu\left[(p-1)_2\mu^2+2(p-1)q\mu+q_1(1+\tilde{\sigma}_q)\right]=0,
\label{eq:power2_GB2}
\\
&& \hs{-10}
p(p+q-3)\mu^2\left[(p-1)_2\mu^2+2(p-1)q\mu+q_1(1+\tilde{\sigma}_q)\right]=0\,.
\label{eq:power2_GB3}
\ena
Then we have either $\mu=0$ or
\bea
(p-1)_2\mu^2+2(p-1)q\mu+q_1(1+\tilde{\sigma}_q)=0\,.
\label{mu_power2_GB}
\ena

For $\mu=0$, inserting it into Eq.~(\ref{eq:power2_GB1}) gives
$\sigma_q=-1$ and $b_0=1$. It is just the Minkowski spacetime with
Milne-type time slicing found above. When Eq.~(\ref{mu_power2_GB}) is
satisfied, eliminating $(1+\tilde{\sigma}_q)$ in Eq.~(\ref{eq:power2_GB1})
yields the equation for $\mu$:
\bea
&& (p+q-2)\mu^2\left[(p-2)(2pq-3p-q+3)\mu^2 \right.\nn
&& \hs{20} \left. +4q(pq-2p-q+3)\mu+2q^2(q-3)\right]=0\,.
\ena
We find the solution
\bea
&& \hs{-10} \mu ={q\over (p-2)(2pq-3p-q+3)}\left[-2(pq-2p-q+3)
\pm\sqrt{2p(q-1)(p+q-3)}\right],
\label{sol:mu_power2_GB}
\\
&& \hs{-10} 1+\tilde{\sigma}_q = \frac{(p-1)q \left\{
6+2p(q-2)-2q \mp \sqrt{2p(q-1)(p+q-3)}\right\} }{(p-2)(q-1)(2pq-3p-q+3)^2}\nn
&& \hs{10} \times \left\{2p(q-1)\pm \sqrt{2p(q-1)(p+q-3)}\right\}.
\label{sol:sigma_power2_GB}
\ena

For $p=3$ and $q=6$, we obtain
$\mu=\frac{3}{2}(-3\pm \sqrt{5})$ and $\tilde{\sigma}_q
=\frac{1}{10}(5 \mp 3\sqrt{5})$, i.e.
$(\mu, \nu) \approx (-1.1459, 1), \s_q=-1, b_0\approx 2.41953$,
or $(\mu, \nu) \approx (-7.8541, 1), \s_q=+1, b_0\approx 0.924176$.
In both cases, the external space is contracting.

For the case of $\sigma_p\neq 0$ and $\sigma_q=0$,
exchanging $\mu, p$ and $\nu, q$, we obtain the solutions.
For $p=3$ and $q=6$, we find $\mu=1$,
$\nu=0, -1$ and $\tilde{\sigma}_p=-1, \frac23$, i.e.
$(\mu, \nu) = (1, 0), \s_p=-1, a_0=1$,
or  $(\mu, \nu) = (1, -1), \s_p=+1, a_0=\frac{2}{3}$.
The first is an exact solution similar to that found for $\s_p=0,\s_q=-1$.
Here the external space is expanding while the internal space is
static or contracting and these are interesting solutions.

\subsection{$\sigma_p \sigma_q\neq 0$}
\label{sec3.3}

\subsubsection{Generalized de Sitter solutions($\epsilon=0$)}

If $\mu=\nu=0$, our basic equations reduce to
\bea
&& \alpha_1 [p_1\tilde \sigma_p + q_1\tilde{\sigma}_q]
+\alpha_2[p_3\tilde\sigma_p^2 + 2p_1q_1\tilde\sigma_p\tilde{\sigma}_q
+q_3\tilde{\sigma}_q^2] = 0, \nn
&& \alpha_1[(p-1)_2\tilde\sigma_p + q_1\tilde{\sigma}_q]
+\alpha_2[(p-1)_4\tilde\sigma_p^2 + 2(p-1)_2q_1\tilde\sigma_p
\tilde{\sigma}_q + q_3\tilde{\sigma}_q^2]=0, \nn
&& \alpha_1 [p_1\tilde\sigma_p + (q-1)_2\tilde{\sigma}_q]
+ \alpha_2[p_3\tilde\sigma_p^2 + 2p_1(q-1)_2\tilde\sigma_p
\tilde{\sigma}_q+(q-1)_4\tilde{\sigma}_q^2]=0.
\ena
It is easy to see that there is no consistent solution.

If either $\mu= 0$ or $\nu= 0$ and the other is nonzero, it is clear that
there is no exact solution. For asymptotic solutions, we can search for
them by setting $A_p=\tilde\sigma_p, A_q=\nu^2$ and $X=0,Y=\nu^2$ for the
first case. We find that there is no real asymptotic solution.
The second case is similar.
For $\mu \nu\neq 0$, if our ansatz for solutions is imposed, it is easy
to see that there is no asymptotic solution if $\mu$ and $\nu$ are
of the opposite signs. If they are of the same sign, either $t\to +\infty$
or $t\to -\infty$ gives $A_p \to \mu^2, A_q \to \nu^2$ and there may be
solutions. However, we find that there is no solution for
Eqs.~(\ref{H_eq}) and (\ref{dynamical_eq1}).
To study time evolution of the system, we need again a numerical analysis.

\subsubsection{Power-law solutions ($\epsilon=1$)}
\label{sec:gbp}

In this case, we first consider the cases when both $\mu$ and $\nu$ are
not equal to 1.

\noindent
{\bf (1) $\mu>1$ and $\nu>1$:}

As $t\to \infty$ with EH dominance, we obtain the asymptotic solutions in
subsection~\ref{gbe1}. However no solutions satisfy the
condition of $\mu>1$ and $\nu>1$ (see Eq. (\ref{Kasner})). This implies
that there is no asymptotic solution of our form.
As $t\to -\infty$, the GB terms become dominant,
and we find no consistent solution.

\noindent
{\bf (2) $\mu<1$ and $\nu<1$:}

As $t\to \infty$ with EH dominance, we again find no consistent solution.
As $t\to -\infty$ with GB dominance, we obtain the asymptotic
solution in subsection~\ref{gbe1}. For the heterotic strings, we find
only one consistent solution, which is $(\mu,\nu)=(-0.19506, 0.59753)$.

\noindent
{\bf (3) $\mu>1$ and $\nu<1$:}

As $t\to \infty$, $A_p\to \mu^2$ and
$A_q\to \tilde{\sigma}_q
e^{2(1-\nu)t}$. This is similar to the case (2)
in the previous section. Then there is no asymptotic solution of our form.
As $t\to -\infty$, $A_p\to \tilde{\sigma}_p e^{2(1-\mu)t}$ and
$A_q\to \nu^2$. We find no solution.

\noindent
{\bf (4) $\mu<1$ and $\nu>1$:}

Here we reach the same result by exchanging $p,\mu$ and $q,\nu$.
No asymptotic solution of our form is obtained.

Next, we discuss the cases in which one of $\mu$ or $\nu$ is equal to
1 and the other is not:

\noindent
{\bf (5) $\mu>1$ and $\nu=1$:}

As $t\to \infty$ with EH dominance, $A_p\to \mu^2$ and $A_q=1+\tilde\s_q$,
and we recover the case of $\sigma_p=0, \sigma_q\neq 0$.
However, there is no solution with $\mu>1$. We do not have any asymptotic
solution of our form. As $t\to -\infty$, $A_p\to
\tilde{\sigma}_p e^{2(1-\mu)t}$ and $A_q=1+\tilde\s_q$.  We again do not
have any asymptotic solution of our form.

\noindent
{\bf (6) $\mu<1$ and $\nu=1$:}

As $t\to \infty$, $A_p$ diverges as $\tilde{\sigma}_p
e^{2(1-\mu)t}$. There is no asymptotic solution of our form.
As $t\to -\infty$, we again recover the case of $\sigma_p=0,
\sigma_q\neq 0$ with the GB-term dominance, namely
$\mu=0,\nu=1,\s_q=-1$, and \p{sol:mu_power2_GB} with \p{sol:sigma_power2_GB}.
Since these asymptotic solutions are consistent with $\mu<1$,
we have asymptotic power-law solutions.
(Note that the first one was an exact solution for $\s_p=0$, but here we are
considering $\s_p\neq 0$.)

\noindent
{\bf (7) $\mu=1$ and $\nu>1$:}

The analysis is almost the same as the case (5).
There is no asymptotic solutions.

\noindent
{\bf (8) $\mu=1$ and $\nu<1$:}

The analysis is almost the same as the case (6),
then we find the asymptotic solutions as $t\to -\infty$, which are
the same as the case of $\sigma_p\neq 0,  \sigma_q= 0$.
We have $\mu=1,\nu=0, \s_p=-1$ and $\mu=1, \nu=-1, \s_p=+1, a_0=\frac{2}{3}$.

Finally, we consider the remaining case.\\
{\bf (9) $\mu=1$ and $\nu=1$:}

Here we have constant $A_p=1+\tilde{\sigma}_p$ and $A_q=1+\tilde{\sigma}_q$.
As $t \to +\infty$, the EH term is dominant, and we have
\bea
&& p_1 A_p +q_1 A_q +2pq=0, \nn
&& (p-1)_2 A_p +q_1 A_q +2(p-1)q=0, \nn
&& p_1 A_p +(q-1)_2 A_q +2p(q-1)=0.
\ena
The solution is given by
\bea
A_p = -\frac{q}{p-1}, \quad
A_q = -\frac{p}{q-1}.
\label{mn1}
\ena
This is the solution found in Ref.~\cite{cosm3} which exhibits eternal
accelerating expansion when higher order effects are taken into account.
For $p=3,q=6$, we have $\tilde\s_p=-4, \tilde\s_q=-\frac{8}{5}$.

For $t\to -\infty$, GB terms are dominant and we get
two independent equations
\bea
&&
p_3 A_p^2+p_1q_1A_pA_q+3p_2qA_p+pq_2A_q+2p_1q_1=0, \\
&&q_3 A_q^2+p_1q_1A_pA_q+p_2qA_p+3pq_2A_q+2p_1q_1=0\,,
\ena
For $p=3$ and $q=6$, we have only one real solution
$A_p=-1.36156, A_q=-1.85305$, i.e. $\sigma_p=\sigma_q=-1$ and
$a_0=0.65073, b_0=0.592032$.

\section{Solutions in M and type II theories}

The higher order corrections to M and type II theories do not involve
GB terms, so we have to take the fourth-order corrections into
account.
{}From our ansatz for solutions, we have
\bea
&&
F_1+F_4+F_S=0,
\label{M:basic1}
\\
&&
F_1^{(p)}+F_4^{(p)}+F_S^{(p)}=0,
\label{M:basic2}
\\
&&
F_1^{(q)}+F_4^{(q)}+F_S^{(q)}=0,
\label{M:basic3}
\,,
\ena
where
\bea
&&F_1=F_1(t, \epsilon,\mu,\nu,A_p,A_q), \nn
&&F_4=F_4(t, \epsilon,\mu,\nu,A_p,A_q), \nn
&&F_S=F_S(t, \epsilon,\mu,\nu,A_p,A_q), \nn
&&F_1^{(p)}=f_1^{(p)}(t, \epsilon,\mu,\nu,A_p,A_q)+X g_1^{(p)}(t,
\epsilon,\mu,\nu,A_p,A_q)+Y h_1^{(p)}(t, \epsilon,\mu,\nu,A_p,A_q), \nn
&&F_4^{(p)}=f_4^{(p)}(t, \epsilon,\mu,\nu,A_p,A_q)+X g_4^{(p)}(t,
\epsilon,\mu,\nu,A_p,A_q)+Y h_4^{(p)}(t, \epsilon,\mu,\nu,A_p,A_q), \nn
&&F_S^{(p)}=F_S^{(p)}(t, \epsilon,\mu,\nu,A_p,A_q), \nn
 &&F_1^{(q)}=f_1^{(q)}(t, \epsilon,\mu,\nu,A_p,A_q)+Y g_1^{(q)}(t,
\epsilon,\mu,\nu,A_p,A_q)+X h_1^{(q)}(t, \epsilon,\mu,\nu,A_p,A_q),\nn
 &&F_4^{(q)}=f_4^{(q)}(t,
\epsilon,\mu,\nu,A_p,A_q)+Y g_4^{(q)}(t, \epsilon,\mu,\nu,A_p,A_q)
+X h_4^{(q)}(t,\epsilon,\mu,\nu,A_p,A_q), \nn
&&F_S^{(q)}=F_S^{(q)}(t, \epsilon,\mu,\nu,A_p,A_q)
\,,
\ena
whose explicit expressions are given in Appendix~B.

\subsection{$\sigma_p = \sigma_q=0$}
\label{sec4.1}

In this case,
$A_p=\mu^2, A_q=\nu^2$ are constants.
We shall discuss the cases of $\epsilon=0$ and  $\epsilon=1$ in order.

\subsubsection{Generalized de Sitter solutions ($\epsilon=0$)}

{}From Appendix~C, we have two algebraic equations:
\bea
&&F_1+F_4+F_S=0
\label{M_eq:cosntraint}
\\
&&H_1+H_4+H_S=0\,,
\label{M_eq:dynamical}
\ena
where $F_1, F_4, F_S, H_1, H_4$, and $H_S$ are functions with respect to
$\mu$ and $\nu$ given in Appendix C. In what follows, we set
$p=3$. The explicit forms of equations
are
\bea
\label{1}
&& \hs{-7}
\alpha_1\left[6 \mu^2 + 6q\mu\nu + q_1 \nu^2\right] + \a_4 q_4 \nu^5
\Big[ 336 \mu^3 + 168 (q-5) \mu^2 \nu + 24(q-5)_6 \mu \nu^2
+(q-5)_7 \nu^3 \Big]  \nn
&& \hs{-5}
-21 \c \Big[ 24 \mu^8 + 2q (\mu^2+\nu^2+\mu\nu )^2 \mu^2 \nu^2+(q+1)_1 \nu^8
+2q(2\mu+\nu)^2 \mu^4 \nu^2  +q_1(\mu+2\nu)^2 \mu^2 \nu^4 \Big] \nn
&& + 24\c ~(3\mu+q\nu) \Big[ 6 \mu^7 + q_1 \nu^7
+ q(\mu+\nu)\mu^2 \nu^2 (\mu^2+\nu^2+\mu\nu ) \Big] =0, \\
\label{2}
&& \hs{-7}
(\mu-\nu)\Big[ \a_1 + 4 \a_4 \left\{ 30(q-1)_4 \mu^2\nu^4
+ 12(q-1)_5 \mu\nu^5 + (q-1)_6 \nu^6 \right\} \nn
&&
+ 2\gamma \left\{ 12\mu^6 - 6(2q-1) \mu^5 \nu +3(q-1)\mu^4 \nu^2 \right.
- (q^2-15q+6) \mu^3\nu^3 -(2q^2-7q-3)\mu^2 \nu^4\nn
&& \hs{10} \left. - (q-1)(q+12) \mu\nu^5 +6(q-1)\nu^6 \right\} \Big]=0.
\ena

Setting $\alpha_1=1$, we have solved these equations numerically.
Before giving the solution, we note on the unit used in our solutions
when the coupling constants $\alpha_4$ and $\c$ are free.
If $\c$ does not vanish, rescaling $\a_4$, $\c$, $\mu$ and $\nu$ as
\bea
\tilde\a_4=\a_4 /|\c| \, \,,\;
\tilde{\c}=\c /|\c| ~(=1 ~{\rm or}~ -1)  \, \,,\;
\tilde{\mu}= \mu |\c|^{1/6}  \,, \;
\mbox{ and }\,\,
\tilde{\nu}= \nu |\c|^{1/6}\,,
\label{nonzeroc}
\ena
we can always set $\c$ to $-1$ if it is negative (or 1 if positive).
We also have to rescale time coordinate as $\tilde{t}=|\c|^{-1/6} t$.
The typical dynamical time scale is then given by $|\c|^{1/6}\sim
O(m_{D}^{-1})$, where $m_{D}=\kappa_{D}^{-2/(D-2)}$ is
the fundamental Planck scale. In particular, for M-theory, we find
$|\c|^{1/6}=6^{-1/6}(4\pi)^{-5/9} m_{11}^{-1}\sim 0.1818176
m_{11}^{-1}$ from Eq.~\p{m}.
After this scaling, we have only one
free parameter $\tilde\a_4$.

If $\gamma=0$ and $\a_4\neq 0$, we can always set $\a_4$ to $-1$ if it is
negative (or 1 if positive), by rescaling $\a_4$, $\mu$ and $\nu$ as
\bea
\bar\a_4=\a_4 /|\a_4| ~(=1 ~~{\rm or}~~-1)\, \,,\;
\bar{\mu}= \mu |\a_4|^{1/6}  \,, \;
\mbox{ and }\,\,
\bar{\nu}= \nu |\a_4|^{1/6}\,.
\label{zeroc}
\ena

Let us now present our results for M-theory and type II superstrings, in
which $\alpha_4$ and $\gamma$ are given by Eq.~\p{m}.
In this paper, we use the above unit as in our previous paper~\cite{MO}.
(We have slightly changed our convention so the numerical results
also a little change from those in \cite{MO}.)
For brevity, we omit a tilde for variables except for
$\tilde\a_4$ and $\tilde\c$.

\noindent
{\bf (1)  M-theory}

For the M-theory, we have
\bea
\tilde\a_4 = - \frac{1}{3 \times 2^5}, \quad
\tilde\c = -1.
\label{m1}
\ena
We then find three solutions
\bea
\label{oursol}
(\mu,\nu) &=& (0.40731,0.40731),~
(0.79683,0.10793),~
(0.55570,0.34253),
\ena
and the time-reversed ones $(-\mu,-\nu)$.

\noindent
{\bf (2)  Type II superstrings}

In type II superstrings, the coefficients $\tilde\a_4$ and $\tilde\c$ are
same as the M-theory, but there are additional terms in the curvature as
well as dilaton~\cite{TBB}. However, we examine what happens
if we simply consider the above theory for ten dimensions ($q=6$).
Since the basic features of the obtained results are the same,
we simply give the solutions. The same remark applies to the following
discussions on type II superstrings.

With the couplings~\p{m1}, we find three solutions
\bea
\label{oursols}
(\mu,\nu) &=& (0.50754, 0.50754),~
 (0.79988, 0.12991),~
 (0.49618, 0.51313),
\ena
and the time-reversed ones.
We thus find that the solutions are qualitatively similar to those
in M-theory.

\subsubsection{Power-law solutions ($\epsilon=1$)}
\label{4thp}

As $t \to \infty$ with EH dominance, we get the same results~\p{Kasner1}
in Sec.~\ref{gbe1}. For $p=3,q=7$, we get
$(\mu,\nu)=(\frac{1\pm \sqrt{21}}{10},\frac{7\mp 3\sqrt{21}}{70})$.

As $t \to -\infty$, the fourth-order terms
dominate. So let us briefly discuss asymptotic power-law solutions only
with quartic terms. Assuming the metric~\p{ansatz} with $\e=1$, we obtain
three algebraic equations:
\bea
&&
\a_4 q_4 \nu^5 \Big[ 336 \mu^3 + 168(q-5) \mu^2 \nu
+ 24 (q-5)_6 \mu\nu^2 + (q-5)_7 \nu^3 \Big] \nn
&&
-\; 7\c \Big[6 \mu^4 (\mu-1)^2 (3 \mu-1)^2 + q_1 \nu^4 (\nu-1)^2
(3\nu -1)^2 + 18 \mu^8 + 3 q_2 \nu^8 \nn
&& \hs{5}
+\; 6q(2\mu+\nu)^2 \mu^4 \nu^2
+ 3q_1(\mu+2\nu)^2 \mu^2 \nu^4 + 6 q \mu^2 \nu^2
\Big( (\mu+\nu-1)^2-\mu \nu \Big)^2 \Big] \nn
&&
+\; \c \Big[ 24 \mu^4(\mu-1)(2 \mu-1)(3 \mu -1)
+ 4q_1 \nu^4(\nu-1)(2\nu-1)(3\nu-1) \nn
&& \hs{10}
+\; 24 q \mu^2 \nu^2 (\mu+\nu-1) \Big( (\mu+\nu-1)^2 - \mu \nu \Big) \Big]
(3 \mu + q\nu -7) =0, \\
&&
4 \a_4 q_4 \nu^5 \Big[ 14 (q-5) \mu^2 \nu + 4(q-5)_6
\mu \nu^2 +4 \mu (\mu-1) \Big(6\mu +(q-5)\nu \Big) \nn
&& \hs{5}
+\; 2(\nu-1) \Big( 30 \mu^2 +12 (q-5) \mu \nu +(q-5)_6 \nu^2 \Big)
\Big] \nn &&
+\; \c \Big[ 6 \mu^4 (\mu-1)^2 (3 \mu-1)^2 + q_1 \nu^4 (\nu-1)^2
(3\nu -1)^2 + 18 \mu^8 + 3 q_2 \nu^8 \nn
&& \hs{10}
+\; 6q (2\mu+\nu)^2 \mu^4 \nu^2
+ 3q_1(\mu+2\nu)^2 \mu^2 \nu^4
+ 6 q \mu^2 \nu^2 \Big( (\mu+\nu-1)^2-\mu \nu \Big)^2 \Big] \nn
&&
-\; \c \Big[ 4 \mu^3(\mu-1)(2 \mu-1)(3 \mu -1)^2
+ 4 \mu^3 (\mu-1)^2 (3 \mu-1)(6 \mu -1)  \nn
&& \hs{10}
+\; 48 \mu ^7 + 8q \mu^3 \nu^2 (2\mu+\nu)(3\mu+\nu)
+ 4q_1 \mu \nu^4(\mu+\nu)(\mu+2\nu) \nn
&& \hs{10}
+\; 4 q \mu \nu^2 \Big( (\mu+\nu-1)(3\mu+\nu-1)-2 \mu \nu\Big)
\Big( (\mu+\nu-1)^2 - \mu \nu \Big) \Big] (3 \mu + q\nu -7) \nn
&&
+\; \c \Big[ 8 \mu^3(\mu-1)(2 \mu-1)(3 \mu -1)
+ 4 q \mu \nu^2 (\mu+\nu-1) \Big( (\mu+\nu-1)^2 - \mu \nu \Big) \Big] \nn
&& \hs{20}
\times (3 \mu + q\nu -7)^2 =0, \\
&&
\a_4 (q-1)_4 \nu^4 \Big[ (q-5)_8 \nu^4 + 24(q-5)_7 \mu\nu^3
+168 (q-5)_6 \mu^2 \nu^2 + 336 (q-5) \mu^3 \nu \nn
&& \hs{5}
+\; 24\mu (\mu-1) \Big( 30 \mu^2 + 12 (q-5) \mu \nu + (q-5)_6\nu^2 \Big) \nn
&& \hs{5}
+\; 8(\nu-1)\Big( 120 \mu^3 +90 (q-5) \mu^2 \nu +18(q-5)_6 \mu\nu^2
+(q-5)_7 \nu^3 \Big) \Big] \nn
&&
+\; \c \Big[ 6 \mu^4 (\mu-1)^2 (3 \mu-1)^2 + q_1 \nu^4 (\nu-1)^2 (3\nu -1)^2
+ 18 \mu^8 + 3 q_2 \nu^8 \nn
&& \hs{10}
+\; 6q (2\mu+\nu)^2 \mu^4 \nu^2 + 3q_1(\mu+2\nu)^2 \mu^2 \nu^4
+ 6 q \mu^2 \nu^2 \Big( (\mu+\nu-1)^2-\mu \nu \Big)^2 \Big] \nn
&&
-\; 2 \c \nu \Big[ (q-1) \nu^2 (\nu-1)(2 \nu-1)(3 \nu -1)^2
+ (q-1) \nu^2 (\nu-1)^2 (3 \nu-1)(6 \nu -1)  \nn
&& \hs{10}
+\; 12(q-1)_2 \nu^6 + 12 \mu^4 (\mu+\nu)(2\mu+\nu)
+ 6(q-1) \mu^2 \nu^2(\mu+2\nu)(\mu+3\nu) \nn
&& \hs{10}
+\; 6 \mu^2 \Big( (\mu+\nu-1)(\mu+3\nu-1)-2 \mu \nu\Big)
\Big( (\mu+\nu-1)^2 - \mu \nu \Big) \Big] (3 \mu + q\nu -7) \nn
&&
+\; 4 \c \nu \Big[(q-1) \nu^2 (\nu-1)(2 \nu-1)(3 \nu -1)
+ 3 \mu^2 (\mu+\nu-1) \Big( (\mu+\nu-1)^2 - \mu \nu \Big) \Big] \nn
&& \hs{20}
\times (3 \mu + q\nu -7)^2 =0.
\ena

Using the values for $\tilde\a_4$ and $\tilde\c$ in Eq.~\p{m1},
we have solved these equations numerically and found
the following four solutions:
\bea
q=7 &&(\mbox{M-theory})\nn
(\mu, \nu) &=&
(0.87610, 0.62453), \;\;
(0.53167, 0.77214), \nn
&&
(0.32052, 0.000168), \;\;
(-0.000877, 0.28898)\,,
\label{mps0}
\\
q=6 &&({\rm Type ~II~ superstrings})\nn
(\mu, \nu) &=&
(5.74269, 5.74269), \;\;
(0.32052, 0.000168), \nn
&&
(0.28829, 0.28829), \;\;
(0.00133, 0.295437).
\label{2ps0}
\ena

\subsection{$\sigma_p =0, \sigma_q\neq 0$
(or $\sigma_p\neq 0, \sigma_q = 0$)}

\subsubsection{Generalized de Sitter solutions ($\epsilon=0$)}

Here we have $A_p=\mu^2, A_q=\nu^2 +\tilde \s_q e^{-2\nu t}, X=\mu^2$
and $Y=\nu^2$. It is easy to see that there is no exact solution unless
$\nu=0$, in which case we have constant $A_p=X=\mu^2, A_q=\tilde \s_q$
and $Y=0$. Our basic equations~\p{M:basic1} and \p{M:basic3}
now give
\bea
\label{M:basic11}
&& \a_1 \left[ p_1 \mu^2+q_1 \tilde \s_q \right]
+\a_4 \Big[ p_7 \mu^8 + 4 p_5 q_1 \mu^6 \tilde \s_q +6 p_3 q_3 \mu^4
\tilde \s_q^2 \nn
&& \hs{30} +4 p_1 q_5 \mu^2 \tilde \s_q^3 + q_7 \tilde \s_q^4 \Big]
+ 3\c \left[ (p-7) p_1 \mu^8 + q_2 \tilde \s_q^4 \right] =0, \\
&& \a_1 \left[ (p+1)_0 \mu^2+(q-1)_2 \tilde \s_q \right]
+\a_4 \Big[ (p+1)_6 \mu^8 + 4 (p+1)_4 (q-1)_2 \mu^6 \tilde \s_q \nn
&& \hs{10}+ 6 (p+1)_2 (q-1)_4 \mu^4 \tilde \s_q^2 +4 (p+1)_0 (q-1)_6 \mu^2
\tilde \s_q^3 + (q-1)_8 \tilde \s_q^4 \Big] \nn
&& \hs{50}+ 3\c \left[ (p+1)_1 \mu^8 + (q-8)(q-1)_2 \tilde \s_q^4 \right] =0.
\ena
We note that Eq.~\p{M:basic2} gives the same equation as \p{M:basic11}
for $\nu=0$ and need not be taken into account.

For $p=3$, we find the following solutions:
\bea
\label{sp=0sqn01}
q=7 &&(\mbox{M-theory})\nn
(\mu, \tilde\s_q)& =& (\pm 0.65615, 0.28708),~(\pm 0.61935, -0.61904),~
(\pm 0.60255, -0.08823)\,,\;\;\;\; \\
\label{sp=0sqn02}
q=6 &&({\rm Type ~II~ superstrings})\nn
(\mu, \tilde\s_q) &=& (\pm 0.76553, 0.45670),~ (\pm 0.62004,
-0.13097).
\ena

For the case of $\sigma_p\neq 0$ and $\sigma_q=0$,
exchanging $\mu, p$ and $\nu, q$, we obtain the solutions with $\mu=0$ and
\bea
\mbox{M-theory}:&&
(\nu,\tilde \s_p)=(\pm 0.49021, 0.63074), \nn
\mbox{Type II superstrings}:&&
(\nu,\tilde \s_p)=(\pm 0.62007, 0.86033).
\label{sp=0sqn03}
\ena

\subsubsection{Power-law solutions ($\epsilon=1$)}
\label{422}

Here we have $A_p=\mu^2, A_q=\nu^2 +\tilde \s_q e^{2(1-\nu) t}, X=\mu(\mu-1)$
and $Y=\nu(\nu-1)$. We have asymptotic solutions in most cases.

\noindent
{\bf (1) $\nu>1$}

For $t \to \infty$, the EH term dominates and $A_q \to \nu^2$.
The solutions are the same as $\s_p=\s_q=0$ case in Sec.~\ref{gbe1}.
However, there is no solution with $\nu>1$.

For $t \to -\infty$, $A_q \to \tilde \s_q e^{2(1-\nu)t}$ and
there is no solution.

\noindent
{\bf (2) $\nu<1$}

For $t \to \infty$, $A_q \to  \tilde \s_q e^{2(1-\nu)t}$ and there is
no solution.

For $t \to -\infty$, $A_q \to \nu^2$ and the solutions are the same as
$\s_p=\s_q=0$ case.

\noindent
{\bf (3) $\nu=1$}

We have $A_p=\mu^2, A_q =1+\tilde\s_q, X=\mu(\mu-1)$ and $Y=0$.

For $t \to \infty$, the EH term dominates and the solutions are
the same as GB case. We have
\bea
\mu=0, \nu=1, \s_q=-1, b_0=1.
\label{sp1}
\ena
Actually this is an exact solution.

For $t \to -\infty$, fourth-order terms dominate.
Our basic independent equations~\p{M:basic1} and \p{M:basic3} give
\bea
\label{eq11}
&& \a_4 \Big[ p_7 \mu^8 + 4 p_5 q_1 \mu^6 A_q + 6 p_3 q_3 \mu^4 A_q^2
+ 4 p_1 q_5 \mu^2 A_q^3 + q_7 A_q^4
+8 \mu \left\{ p_6 q \mu^6 + 3 p_4 q_2 \mu^4 A_q \right. \nn
&& \left.+ 3 p_2 q_4 \mu^2 A_q^2 + p q_6 A_q^3 \right\}
+ 24 \mu^2 \left\{ p_5 q_1 \mu^4 + 2 p_3 q_3 \mu^2 A_q + p_1 q_5 A_q^2\right\}
+ 32 \mu^3 \left\{ p_4 q_2 \mu^2 + p_2 q_4 A_q \right\} \nn
&&\hs{5} + 16 p_3 q_3 \mu^4 \Big]
+ \c \Big[ -7 \tilde L_4 +(p\mu +q -7)(p\mu M_X+q M_Y)+2(A_q-1)qN_q \Big]
=0, \\
\label{eq21}
&& \tilde f_4^{(q)} + \tilde h_4^{(q)}\mu(\mu-1) + \tilde F_S^{(q)}=0,
\ena
where
\bea
&& \tilde L_4 =p_1 \mu^4(\mu-1)^2 (3\mu-1)^2 + 2pq\mu^4(\mu-1)^2
+3p_2 \mu^8+ 3q_2 A_q^4 \nn
&& \hs{40} +\; p_1 q \mu^4 (2\mu+1)^2 +pq_1 \mu^2(\mu+2A_q)^2,\\
&& M_X = 4\left[(p-1)\mu^3(\mu-1)(2\mu-1)(3\mu-1)+q\mu^3(\mu-1)\right],\\
&& M_Y = 4p\mu^4(\mu-1),\\
&& N_q = 4(q-1)\left[ 3(q-2)A_q^3+p\mu^2(\mu+2A_q)\right], \\
&& \tilde f_4^{(q)} = \a_4 \Big[ p_7 \mu^8 + 4p_5 (q-1)_2 \mu^6 A_q
+ 6 p_3 (q-1)_4 \mu^4 A_q^2 + 4 p_1 (q-1)_6 \mu^2 A_q^3 + (q-1)_8 A_q^4 \nn
&&\hs{10} +8 \mu \left\{ p_6 (q-1) \mu^6 + 3 p_4 (q-1)_3 \mu^4 A_q
+ 3 p_2 (q-1)_5 \mu^2 A_q^2 + p (q-1)_7 A_q^3 \right\}\nn
&& \hs{10}+ 24 \mu^2 \left\{ p_5 (q-1)_2 \mu^4 + 2 p_3 (q-1)_4 \mu^2 A_q
+ p_1 (q-1)_6 A_q^2 \right\} \nn
&& \hs{15} + 32 \mu^3 \left\{ p_4 (q-1)_3 \mu^2 + p_2 (q-1)_5 A_q \right\}
+ 16 p_3 (q-1)_4 \mu^4 \Big],\\
&& \tilde h_4^{(q)} = 8p \a_4 \Big[ (p-1)_6 \mu^6 + 3(p-1)_4 (q-1)_2 \mu^4 A_q
+ 3 (p-1)_2 (q-1)_4 \mu^2 A_q^2 + (q-1)_6 A_q^3 \nn
&&\hs{10} +6 \mu \left\{ (p-1)_5 (q-1) \mu^4 + 2 (p-1)_3 (q-1)_3 \mu^2 A_q
+ (p-1) (q-1)_5 A_q^2 \right\} \nn
&&\hs{10}+12 \mu^2 \left\{ (p-1)_4 (q-1)_2 \mu^2 + (p-1)_2 (q-1)_4 A_q \right\}
+ 8 (p-1)_3 (q-1)_3 \mu^3 \Big],\\
&& \tilde F_S^{(q)} = \c \Big[ \tilde L_4 - (p\mu+q-7) \left\{
M_Y+2N_q+p\mu U \right\} + (p\mu+q-7)^2 M_Y -2(A_q-1) N_q \Big],\nn \\
&& U = 4\left[ \mu^3(\mu-1)^2+(p-1)\mu^3(\mu+1)(2\mu+1)
+(q-1)\mu(\mu+A_q)(\mu+2A_q)\right].
\ena

For $p=3$, we find the solution~\p{sp1} and
\bea
q=7 &&(\mbox{M-theory})\nn
(\mu, \tilde \s_q) &=&  (14.8319, -413.5411),~ (0.7335, -0.3062),
\\[1em]
q=6 &&({\rm Type ~II~ string})\nn
(\mu, \tilde \s_q) &=& (4.0305, 8.7771),~(0.4484, -1.2490),~
(-9.7439, -94.7146).
\ena
(We also find a solution $(\mu,\tilde\s_q)=(0, 7)$ to Eqs.~\p{eq11} and
\p{eq21}, but this is the special case of $\dot u_1=0$, and then
Eq.~\p{M:basic2} must be checked, as discussed in subsection~\ref{sec2.1}.
We find that it is not satisfied asymptotically and this is no a solution.)
First of these gives an interesting inflationary solution.

For the case of $\sigma_p\neq 0$ and $\sigma_q=0$,
exchanging $\mu, p$ and $\nu, q$, we obtain the solutions:
\bea
\mbox{exact solution}: && (\mu,\nu,\s_p) = (1,0,-1), \nn
\mbox{past asymptotic solutions}: && (\mu,\nu,\tilde\s_p)
=(1, 0.7181, -1.5485),~(1, 0.0417, -1.0204), \nn
&& \hs{20}
(1, -14.1607, -138.1063),
\ena
for M-theory and
\bea
\mbox{exact solution}: && (\mu,\nu,\s_p) = (1,0,-1), \nn
\mbox{past asymptotic solutions}: && (\mu,\nu,\tilde\s_p)
=(1, 6.1725, 75.9086),~(1, 0.0358, -1.0173), \nn
&& \hs{20}
(1, -26.8744, -961.1752),
\ena
for type II superstrings.

\subsection{$\sigma_p\sigma_q\neq 0$}

\subsubsection{Generalized de Sitter solutions ($\epsilon=0$)}

If $\mu=\nu=0$, our basic equations reduce to
\bea
&& \alpha_1 [p_1\tilde \s_p + q_1\tilde\s_q]
+\a_4 [p_7 \tilde\s_p^4 + 4p_5q_1\tilde\s_p^3 \tilde\s_q
+ 6p_3 q_3 \tilde\s_p^2 \tilde\s_q^2 + 4p_1q_5\tilde\s_p \tilde\s_q^3
+q_7\tilde\s_q^4] \nn
&& \hs{50} + 3 \c [ p_2 \tilde\s_p^4 + q_2 \tilde\s_q^4] = 0, \nn
&& \a_1[(p-1)_2\tilde\s_p + q_1\tilde\s_q]
+\a_4 [(p-1)_8\tilde\s_p^4 + 4(p-1)_6q_1\tilde\s_p^3 \tilde\s_q
+ 6 (p-1)_4q_3\tilde\s_p^2 \tilde\s_q^2 \nn
&&\hs{10} + 4(p-1)_2q_5\tilde\s_p \tilde\s_q^3 + q_7\tilde\s_q^4]
+ 3 \c [ (p-8)(p-1)_2 \tilde\s_p^4 + q_2 \tilde\s_q^4] = 0, \\
&& \a_1[p_1 \tilde\s_p + (q-1)_2 \tilde\s_q]
+\a_4 [p_7\tilde\s_p^4 + 4p_5 (q-1)_2 \tilde\s_p^3 \tilde\s_q
+ 6 p_3(q-1)_4 \tilde\s_p^2 \tilde\s_q^2 \nn
&&\hs{10} + 4p_1(q-1)_6\tilde\s_p \tilde\s_q^3 + (q-1)_8\tilde\s_q^4]
+ 3 \c [ p_2 \tilde\s_p^4 +(q-8)(q-1)_2 \tilde\s_q^4] = 0. \nonumber
\ena

For both M-theory with $p=3,q=7$ and type II theory with $p=3,q=6$,
we find that there is no solution.

If either $\mu= 0$ or $\nu= 0$ and the other is nonzero, it is clear that
there is no exact solution. For asymptotic solutions, we can search for
them by setting $A_p=\tilde\sigma_p, A_q=\nu^2$, $X=0$ and $Y=\nu^2$ for the
first case. The solution is for $t\to +\infty (-\infty)$ for
$\mu \mbox{ or }\nu >0 (<0)$. The basic equations are
\bea
&& \a_1 [ p_1 \tilde\s_p + q_1\nu^2 ]
+\a_4 [p_7 \tilde\s_p^4 +4p_5 q_1 \tilde\s_p^3 \nu^2
+6p_3 q_3 \tilde\s_p^2 \nu^4 +4 p_1 q_5 \tilde\s_p \nu^6 +q_7 \nu^8] \nn
&& \hs{70} + 3\c [ p_2 \tilde\s_p^4 + (q-7)q_1 \nu^8] =0, \nn
&& \a_1[(p-1)_2 \tilde\s_p + (q+1)_0 \nu^2]
+\a_4 [(p-1)_8 \tilde\s_p^4 + 4(p-1)_6 (q+1)_0 \tilde\s_p^3 \nu^2 \nn
&& \hs{20}+ 6(p-1)_4(q+1)_2 \tilde\s_p^2
\nu^4 + 4(p-1)_2(q+1)_4 \tilde\s_p \nu^6 + (q+1)_6 \nu^8] \nn
&& \hs{50} + 3\c [ (p-8)(p-1)_2 \tilde\s_p^4 +(q+1)_1 \nu^8]=0.
\ena

We find for M-theory that there are solutions with $\mu=0$ and
\bea
(\tilde \s_p, \nu) = (0.63074, \pm 0.49021),
\ena
and for type II superstrings
\bea
(\tilde \s_p, \nu) = (0.86033, \pm 0.62007),
\ena
for $t \to +\infty\; (-\infty)$ for $\nu >0\; (< 0)$

The second case is obtained by exchanging $p,\mu$ and $q,\nu$.
The solutions are
\bea
(\tilde \s_q, \mu) = (0.28708, \pm 0.65615),~ (-0.61904, \pm 0.61935),~
(-0.08823,\pm 0.60255),
\ena
for M-theory and
\bea
(\tilde \s_q, \mu) = (0.45670, \pm 0.76553),~ (-0.13097, \pm 0.62004),
\ena
for type II superstrings. They are qualitatively the same.

For $\mu \nu\neq 0$, if our ansatz for solutions is imposed, it is easy
to see that there is no asymptotic solution if $\mu$ and $\nu$ are
of the opposite signs. If they are of the same sign, either $t\to +\infty$
or $t\to -\infty$ gives $A_p \to \mu^2, A_q \to \nu^2$ and there may be
solutions. This implies that inflationary solutions with positive
eigenvalues are obtained for asymptotic infinite future, so that these are
not interesting from the cosmological point of view. However it may be
useful to check if there are any solutions of this type.
In fact we find that there are asymptotic solutions for M-theory
\bea
(\mu,\nu) = \pm(0.79683, 0.10792),~ \pm(0.55570, 0.34253),~
\pm(0.40731,0.40731),
\ena
where negative (positive) one is for $t\to -\infty \;(\infty)$.
For type II superstrings, we have
\bea
(\mu,\nu) = \pm(0.79988, 0.12991),~ \pm(0.50754, 0.50754),~
\pm(0.49618, 0.51313).
\ena

\subsubsection{Power-law solutions ($\epsilon=1$)}

In this case, we first consider the cases when both $\mu$ and $\nu$ are
not equal to 1.

\noindent
{\bf (1) $\mu>1$ and $\nu>1$:}

For $t\to \infty$, the EH term dominates and we obtain
the asymptotic solutions in subsection~\ref{gbe1}. Again no solutions
satisfy the condition of $\mu>1$ and $\nu>1$ (see Eq. (\ref{Kasner}))
and hence there is no asymptotic solution of our form.

As $t\to -\infty$, the fourth-order terms become dominant and
we find no consistent solution from the forth-order terms.

\noindent
{\bf (2) $\mu<1$ and $\nu<1$:}

As $t\to \infty$ with EH dominance, we again find no consistent solution.
As $t\to -\infty$ with fourth-order-term dominance, we obtain the asymptotic
solutions in Eqs.~\p{mps0} and \p{2ps0} in subsection~\ref{4thp}.

\noindent
{\bf (3) $\mu>1$ and $\nu<1$:}

As $t\to \infty$, $A_p\to \mu^2$ and $A_q\to \tilde{\sigma}_q
e^{2(1-\nu)t}$. This is similar to the case (2) in subsection~\ref{422}.
There is no asymptotic solution of our form.
As $t\to -\infty$, $A_p\to \tilde{\sigma}_p e^{2(1-\mu)t}$ and
$A_q\to \nu^2$. We find no solution.

\noindent
{\bf (4) $\mu<1$ and $\nu>1$:}

Here we reach the same result by exchanging $p,\mu$ and $q,\nu$.
No asymptotic solution of our form is obtained.

Next, we discuss the cases in which one of $\mu$ or $\nu$ is equal to
1 and the other is not:

\noindent
{\bf (5) $\mu>1$ and $\nu=1$:}

As $t\to \infty$ with EH dominance, $A_p\to \mu^2$,
and we recover the case of $\sigma_p=0, \sigma_q\neq 0$.
However, there is no solution with $\mu>1$.
We do not have any asymptotic solution of our form.
As $t\to -\infty$ with fourth-order-term dominance,
$A_p\to\tilde\s_p e^{2(1-\mu)t}$.
We again do not have any asymptotic solution of our form.

\noindent
{\bf (6) $\mu<1$ and $\nu=1$:}

As $t\to \infty$ with EH dominance, $A_p$ diverges as $\tilde{\sigma}_p
e^{2(1-\mu)t}$. There is no asymptotic solution of our form.
As $t\to -\infty$, we again recover the case of $\sigma_p=0,
\sigma_q\neq 0$ with the fourth-order-term dominance~\p{sp1}.
Since this asymptotic solution is consistent with $\mu<1$,
we have an asymptotic power-law solution. (Note that this was
an exact solution for $\s_p=0$.)

\noindent
{\bf (7) $\mu=1$ and $\nu>1$:}

The analysis is almost the same as the case (5).
There is no asymptotic solution.

\noindent
{\bf (8) $\mu=1$ and $\nu<1$:}

The analysis is almost the same as the case (6),
then we find the asymptotic solution as $t\to -\infty$, which is
the same as the case of $\sigma_p\neq 0,  \sigma_q= 0$.

Finally, we consider the remaining case.\\
{\bf (9) $\mu=1$ and $\nu=1$:}

Here we have constant $A_p=1+\tilde{\sigma}_p$ and $A_q=1+\tilde{\sigma}_q$.
As $t \to +\infty$ with EH dominance, and we recover the solution~\p{mn1}
of subsection~\ref{sec:gbp}. For $p=3,q=7$, we get
$(\tilde\s_p,\tilde\s_q)=(-\frac92, -\frac32)$.

For $t\to -\infty$ with fourth-order-term dominance, we get
two independent equations for $A_p=1+\tilde\s_p, A_q=1+\tilde\s_q$:
\bea
&&
\a_4 \Big[ p_7 A_p^4 + 4 p_5 q_1 A_p^3 A_q + 6 p_3 q_3 A_p^2 A_q^2
+ 4 p_1 q_5 A_p A_q^3 + q_7 A_q^4
+8( p_6 q A_p^3 + 3 p_4 q_2 A_p^2 A_q \nn
&& + 3 p_2 q_4 A_p A_q^2 + p q_6 A_q^3 )
+ 24( p_5 q_1 A_p^2 + 2 p_3 q_3 A_p A_q + p_1 q_5 A_q^2)
+ 32( p_4 q_2 A_p + p_2 q_4 A_q) \nn
&& \hs{10} + 16 p_3 q_3 \Big]
+ \c \left[ -7\tilde L_4 + 2p (A_p-1) N_p + 2q (A_q-1) N_q \right]=0, \\
&& \a_4 \Big[ (p-1)_8 A_p^4 + 4 (p-1)_6 q_1 A_p^3 A_q + 6 (p-1)_4 q_3
A_p^2 A_q^2 + 4 (p-1)_2 q_5 A_p A_q^3 + q_7 A_q^4 \nn
&& + 8 \left\{ (p-1)_7 q A_p^3 + 3 (p-1)_5 q_2 A_p^2 A_q
+ 3 (p-1)_3 q_4 A_p A_q^2 + (p-1) q_6 A_q^3 \right\} \nn
&& + 24 \left\{ (p-1)_6 q_1 A_p^2 + 2 (p-1)_4 q_3 A_p A_q
+ (p-1)_2 q_5 A_q^2\right\}
+ 32 \left\{ (p-1)_5 q_2 A_p + (p-1)_3 q_4 A_q\right\} \nn
&& \hs{10} + 16 (p-1)_4 q_3 \Big]
+ \c \left[ \tilde L_4 - (p+q-7)(2 N_p+q U)- 2(A_p-1)N_p \right]=0,
\ena
where
\bea
&& \tilde L_4 = 3 p_2A_p^4 +3q_2 A_q^4 + p_1 q(2A_p+1)^2+ p q_1(2A_q+1)^2, \nn
&& N_p = 4(p-1) \left[ 3(p-2)A_p^3 + q(2 A_p+1) \right], \nn
&& N_q = 4(q-1) \left[ 3(q-2)A_q^3 + p(2 A_q+1) \right], \nn
&& U = 4 \left[ (p-1)(A_p+1)(2A_p+1) + (q-1)(A_q+1)(2A_q+1) \right].
\ena

For $p=3$, we have four real solutions
\bea
q=7 &&(\mbox{M-theory})\nn
(A_p, A_q) &=& (12.2143, -10.4313), ~ (0.43403, -0.60288),\nn
&& (0.19127,0.73878),~(-2.10241, -0.19306)\\[1em]
q=6 &&({\rm Type ~II~string})\nn
(A_p, A_q) &=& (5.5316, 3.1354), ~ (0.21214, -0.66202),\nn
&& (-1.3472, -0.31116),~ (-33.5609, 19.4154).
\ena

\section{Summary and discussions}

We have found generalized de Sitter solutions
\bea
a\propto e^{\mu t}, ~~b\propto e^{\nu t}
~~~{\rm for}~\epsilon=0,
\label{gdS}
\ena
and power-law solutions
\bea
a\propto \tau^{\mu},  ~~ b\propto
\tau^{\nu} ~~~~{\rm for}~ \epsilon=1.
\label{power1}
\ena
In the Einstein frame,
\bea
a_E\propto t_E^\lambda ~~, ~~~\phi\sim
\phi^{(0)}+\phi_1\ln [t_E/t_E^{(0)}]\,,
\label{power2}
\ena
where
\bea
&&\lambda=1+{(p-1)\mu\over q\nu}~~,~~~
\phi_1={(p-1)\nu\over q} ~~~~~~~~{\rm for}~\epsilon=0,
\nonumber
\\
&&\lambda={(p-1)\mu+q\nu\over (p-1)+q\nu}~,~~
\phi_1={(p-1)\nu\over (p-1)+q\nu}~~~{\rm for}~\epsilon=1.
\ena

Note that the values of $\mu$ and $\nu$ in generalized de Sitter solutions
(\ref{gdS}) depend on the choice of the unit. In the heterotic string theories,
we adopt $\alpha'=1$, while in the M-theory and the type II string theory,
we use the unit of $|\gamma|=1$, i.e.
$m_{11}=6^{-1/2}(4\pi)^{-5/9}\sim 0.1818176$.
If we set $m_{11}=1$, the values of $\mu$ and $\nu$ in the following
tables should be multiplied by the factor
$6^{1/2}(4\pi)^{5/9}\sim 5.5$.
On the other hand, the power exponent $\mu$ and $\nu$ in the power-law
solutions (\ref{power1}) or $\lambda$ in (\ref{power2}) are dimensionless and
they do not depend on the choice of the unit.

We summarize our results in the following tables for the cases of
the heterotic string theories and M-theory in order.
For asymptotic solutions, the time regions for $t_E$ where the solutions
are valid in the Einstein frame are also included.
In the last lines of the tables, we include the type of two spaces ($ds_p^2,
ds_q^2$). K means the kinetic dominant space, in which the curvature term
($\sigma_p$, or $\sigma_q$) is either zero or can be asymptotically ignored.
M denote the Milne-type space, which is described by
$ds^2=-dt^2+t^2 ds_p^2+\cdots$ with $\sigma_p=-1$, or $ds^2=-dt^2+\cdots+t^2
ds_q^2$ with $\sigma_q=-1$.
Similarly, we define a constant curvature space C by $\sigma_p=1$ or
$\sigma_q=1$, and
S$_0$ and S$_\pm$ are static spaces with zero curvature and positive (or
negative) curvature, respectively.
The result for the type II string model, which is similar to
the case of M-theory, is given in Appendix D.

\subsection{Heterotic strings}

Exact solutions are given in Table~\ref{table_3},
future asymptotic solutions in Table~\ref{table_4}
and past asymptotic solutions in Table~\ref{table_5}, where -- means
that the radius can be arbitrary.

Since we are interested in inflation in string theories,
we pick up such solutions and give comments on those.
In the original frame, we find HE1$_+$ (HF3, HP3)
give an exponential expansion whereas HE2$_+$ (HF4, HP4) and HP5 give
a power-law inflation. In the former solutions the extra space expands
exponentially, while the internal space shrinks exponentially.
However, in the Einstein frame, they correspond to a non-inflationary
power-law expansion and HE2$_+$ (HF4, HP4) and HP5 give a power-law
inflation. Another interesting observation is that we could obtain
an expansion of the universe in the Einstein frame from
an external space shrinking in the original frame
[HE1$_-$ (HF1, HP1), HF6, HP6, HP7, HP8].

\begin{table}[H]
\caption{Heterotic string: exact solutions. K, M, and S$_0$ mean a kinetic
dominance, a Milne type space, and a flat static space, respectively.
-- means that the radius can be arbitrary.}
\begin{center}
{\footnotesize
\begin{tabular}{|c||c||c|c||c|c|c|c||c|c||c|}
\hline
&$\epsilon$&$\sigma_p$&$\sigma_q$&$\mu$&$\nu$&$a_0$&$b_0$&
$\lambda$&$\phi_1$& type \\
\hline
\hline
HE1$_\pm$&0&0&0&$\pm 1.366$&$\mp 0.9657$ & -- & -- & 0.5285&$\mp 0.3219$&K K\\
\hline
HE2$_\pm$&0&0&0&$\pm 2.506$&$\mp 0.3916$&--&--&$-1.132$&$\mp 0.1305$&K K\\
\hline
\hline
HE3&1&0&$-1$&0&1& -- &1&0.75&0.25&S$_0$ M\\
\hline
HE4&1&$-1$&0&1&0&1& -- &1&0&M S$_0$\\
\hline
\end{tabular}
}
\label{table_3}
\end{center}
\end{table}

\begin{table}[H]
\caption{Heterotic string : future asymptotic solutions
($t\rightarrow \infty$).  M means a Milne type space.
The time regions for $t_E$ where the solutions are valid in
the Einstein frame are also included.}
\begin{center}
{\footnotesize
\begin{tabular}{|c||c||c|c||c|c|c|c||c|c|c||c|}
\hline
&$\epsilon$&$\sigma_p$&$\sigma_q$&$\mu$&$\nu$&$a_0$&$b_0$&
$\lambda$&$\phi_1$&$t_E$& type \\
\hline
HF1&0&0&$\pm 1$&$-1.366$&0.9657& -- & -- &0.5285&$0.3219$&$\rightarrow
\infty$&HE1$_-$\\
\hline
HF2&0&0&$\pm 1$&$-2.506$&0.3916& -- & -- &$-1.132$&$0.1305$&$\rightarrow
\infty$&HE2$_-$\\
\hline
HF3&0&$\pm
1$&0&$1.366$&$-0.9657$& -- & -- &0.5285&$-0.3219$&$\sim 0$&HE1$_+$\\
\hline
HF4&0&$\pm 1$&0&$2.506$&$-0.3916$& -- & -- &$-1.132$&$-0.1305$&$\sim 0$
& HE2$_+$ \\
\hline
\hline
HF5&1&0&0&0.5556&$-0.1111$& -- & -- & 0.3333&$-0.1667$&$\rightarrow
\infty$&{\rm Kasner}\\
\hline
HF6&1&0&0&$-0.3333$&$0.3333$& -- & -- &0.3333&0.1667&$\rightarrow \infty$
&{\rm Kasner}\\
\hline
HF7&1&$-1$&$-1$&1&1&0.5&0.7906&1 &$0.25$ &$\rightarrow \infty$&M M\\
\hline
\end{tabular}
}
\label{table_4}
\end{center}
\end{table}

\begin{table}[H]
\caption{Heterotic string: past asymptotic solutions
($t\rightarrow -\infty$). K, M, and C mean a kinetic dominance, a Milne type
space, and a constant curvature space, respectively.}
\begin{center}
{\footnotesize
\begin{tabular}{|c||c||c|c||c|c|c|c||c|c|c||c|}
\hline
&$\epsilon$ & $\sigma_p$&$\sigma_q$&$\mu$&$\nu$&$a_0$&$b_0$&
$\lambda$&$\phi_1$&$t_E$& type \\
\hline
HP1 &0&$\pm 1$&0&$-1.366$&$0.9657$&--&--&0.5285&$0.3219$&$\sim 0$&HE1$_-$\\
\hline
HP2 &0&$\pm 1$&0&$-2.506$&$0.3916$&--&--&$-1.132$&$0.1305$&$\sim 0$&HE2$_-$\\
\hline
HP3 &0&0&$\pm 1$&$1.366$&$-0.9657$&--&--&0.5285&$-0.3219$&$\rightarrow
-\infty$&HE1$_+$\\
\hline
HP4 &0&0&$\pm 1$&$2.506$&$-0.3916$&--&--&$-1.132$&$-0.1305$&$\rightarrow
-\infty$&HE2$_+$\\
\hline
\hline
HP5 &1&0&$0, \pm 1$&1.517&$-0.2585$&--&--&3.3029&$-1.15145$&$\sim 0$&K K\\
\hline
HP6 &1&0,$\pm 1$&0&$-1.877$&1.439&--&--&0.4589&0.2706&$\sim 0$&K K\\
\hline
HP7&1&$0, \pm 1$&$0, \pm 1$&$-0.1951$&0.5975&--&--&0.5720&0.2140&$\sim 0$&K K\\
\hline
HP8 & 1 & $0,\pm 1$ &$-1$&$-1.146$&1&--&2.420&0.4635&0.25&$\sim 0$&K M\\
\hline
HP9 & 1 & $0,\pm 1$ &1&$-7.854$&1&--&0.9242&$-1.214$&0.25&$\sim 0$&K C\\
\hline
HP10&1&$\pm 1$&$-1$&0&1&--&1&0.75&0.25&$\sim 0$&K M\\
\hline
HP11&1&$-1$&$\pm 1$&1&0&1&--&1&0&$\sim 0$&M K\\
\hline
HP12&1&$1$&$0, \pm 1$&1&$-1$&1.2247&--&1&0.5&$\rightarrow -\infty$&C K\\
\hline
HP13&1&$-1$&$-1$&1&1&$0.6507$&0.5920&1&0.25&$\sim 0$&M M\\
\hline
\end{tabular}
}
\label{table_5}
\end{center}
\end{table}

\subsection{M-theory}

Exact solutions are summarized in Table~\ref{table_6},
future asymptotic solutions in Table~\ref{table_7}
and past asymptotic solutions in Table~\ref{table_8}.

\begin{table}[H]
\caption{M-theory: exact solutions. K, S$_\pm$, S$_0$, and M mean
a kinetic dominance, a static space with positive (or negative) curvature,
a flat static space, and a Milne type space. }
\begin{center}
{\footnotesize
\begin{tabular}{|c||c||c|c||c|c|c|c||c|c||c|}
\hline
&$\epsilon$&$\sigma_p$&$\sigma_q$&$\mu$&$\nu$&$a_0$&$b_0$&
$\lambda$&$\phi_1$& type \\
\hline
\hline
ME1$_\pm$&0 & 0 & 0 & $\pm 0.7968$ & $\pm 0.1079$ & -- & -- & 3.1099 & $\pm
0.03083$&K K
\\
\hline
ME2$_\pm$&0 & 0 & 0 & $ \pm 0.5557 $ & $\pm 0.3425$ & -- & -- &1.4636& $\pm
0.09786$&K K
\\
\hline
ME3$_\pm$&0 & 0 & 0 & $\pm 0.4073$ & $\pm 0.4073$ & -- & -- &1.2857 & $\pm
0.1164$&K K\\
\hline
ME4$_\pm$&0 & 0 & 1 & $\pm 0.6562$ & 0 & -- & 1.866 & $e^{\mu t_E}$ &
0&K S$_+$\\
\hline
ME5$_\pm$&0 & 0 & $-1$ & $\pm 0.6194$ & 0 & -- & 1.271 & $e^{\mu t_E}$ & 0
&K S$_-$\\
\hline
ME6$_\pm$&0 & 0 & $-1$ & $\pm 0.6026$ & 0 & -- & 3.367 & $e^{\mu t_E}$ & 0
&K S$_-$\\
\hline
ME7$_\pm$&0 & 1 & 0 & 0 & $\pm 0.4902$ & 1.259 & -- &1& $\pm 0.1401$ &S$_+$ K\\
\hline
\hline
ME8&1 & 0 & $-1$ & 0 & 1 & -- & 1 & 0.7778 &  0.7778 &S$_0$ M\\
\hline
ME9&1 & $-1$ & 0 & 1 & 0 & 1 & -- & 1 &  0 &M S$_0$\\
\hline
\end{tabular}
}
\label{table_6}
\end{center}
\end{table}

\begin{table}[htb]
\caption{M-theory: future asymptotic solutions ($t\to \infty$).
M means a Milne type space.}
\begin{center}
{\footnotesize
\begin{tabular}{|c||c||c|c||c|c|c|c||c|c|c||c|}
\hline
&$\epsilon$&$\sigma_p$&$\sigma_q$&$\mu$&$\nu$&$a_0$&$b_0$&
$\lambda$&$\phi_1$&$t_E$& type \\
\hline
MF1&0 & 1 & $\pm 1$ & 0 & 0.4902 & 1.259 & -- & 1 & 0.14006
&$\rightarrow \infty$&ME7$_+$\\
\hline
MF2&0 & $\pm 1$ & 1 & $0.6562$ & 0 & -- & 1.866 &$e^{\mu t_E}$& 0&$\rightarrow
\infty$&ME4$_+$ \\
\hline
MF3&0 & $\pm 1$ & $-1$ & $0.6194$ & 0 & -- & 1.271 &$e^{\mu t_E}$& 0
&$\rightarrow \infty$&ME5$_+$ \\
\hline
MF4&0 & $\pm 1$ & $-1$ & $0.6026$ & 0 & -- & 3.367 &$e^{\mu t_E}$ & 0
&$\rightarrow \infty$&ME6$_+$ \\
\hline
MF5&0 & $\pm 1$ & $\pm 1$ & $0.7968$ & 0.1079 & -- & -- &  3.1099 & 0.03083
&$\rightarrow \infty$&ME1$_+$ \\
\hline
MF6&0 & $\pm 1$ & $\pm 1$ & $0.5557$ & 0.3425 & -- & -- & 1.4636&  0.09786
&$\rightarrow \infty$&ME2$_+$ \\
\hline
MF7&0 & $\pm 1$ & $\pm 1$ & $0.4073$ & 0.4073 & -- & -- &1.2857 & 0.1164
&$\rightarrow \infty$&ME3$_+$\\
\hline
\hline
MF8&1&0&0&0.5583&$-0.0964$& -- & -- &0.3333 & $-0.1455$ &$\rightarrow \infty$
& Kasner \\
\hline
MF9&1&0&0&$-0.3583$&0.2964& -- & -- &0.3333 & 0.1455 &$\rightarrow \infty$
& Kasner \\
\hline
MF10&1 & $-1$ & $-1$ & 1 & 1 & 0.4714 & 0.8165 & 1 & 0.2222&$\rightarrow
\infty$& M M
\\
\hline
\end{tabular}
}
\label{table_7}
\end{center}
\end{table}

\begin{table}[H]
\caption{M-theory: past asymptotic solutions
 ($t\rightarrow -\infty$). K, S$_\pm$, S$_0$, M  and C mean
a kinetic dominance, a static space with positive (or negative) curvature,
a flat static space,  a Milne type space, and a constant curvature space,
respectively.}
\begin{center}
{\footnotesize
\begin{tabular}{|c||c||c|c||c|c|c|c||c|c|c||c|}
\hline
&$\epsilon$&$\sigma_p$&$\sigma_q$&$\mu$&$\nu$&$a_0$&$b_0$&
$\lambda$&$\phi_1$&$t_E$& type \\
\hline
MP1&0 & 1 & $\pm 1$ & 0 & $-0.4902$ & 1.259 & -- & 1 & $-0.14006$
&$\rightarrow -\infty$&ME7$_-$\\
\hline
MP2&0 & $\pm 1$ & $1$ & $-0.6562$ & 0 & -- & 1.866 &  $e^{\mu t_E}$ & 0
&$\rightarrow -\infty$&
ME4$_-$\\
\hline
MP3&0 & $\pm 1$ & $-1$ & $-0.6194$ & 0 & -- & 1.271 &  $e^{\mu t_E}$ &
0&$\rightarrow -\infty$&ME5$_-$ \\
\hline
MP4&0 & $\pm 1$ & $-1$ & $-0.6026$ & 0 & -- & 3.367 &  $e^{\mu t_E}$ & 0
&$\rightarrow -\infty$&
ME6$_-$\\
\hline
MP5&0 & $\pm 1$ & $\pm 1$ & $-0.7968$ & $-0.1079$ & -- & -- & 3.1099 &
$-0.03083$&$\rightarrow -\infty$&ME1$_-$ \\
\hline
MP6&0 & $\pm 1$ & $\pm 1$ & $-0.5557$ & $-0.3425$ & -- & -- & 1.4636&
$-0.09786  $&$\rightarrow -\infty$& ME2$_-$\\
\hline
MP7&0 & $\pm 1$ & $\pm 1$ & $-0.4073$ & $-0.4073$ & -- & -- &  1.2857 &
$-0.1164$&$\rightarrow -\infty$&ME3$_-$
\\
\hline
\hline
MP8&1 & $0,\pm 1$ & $0,\pm 1$ & 0.87610 & 0.62453 & -- &-- & 0.9611 &
0.1960 &$\sim 0$&K K\\
\hline
MP9&1 & $0,\pm 1$ & $0,\pm 1$ & 0.53167 & 0.77214 &-- & -- & 0.8735&
0.2085&$\sim 0$&K K
\\
\hline
MP10&1 & $0,\pm 1$ & $0,\pm 1$ & 0.32052 & 0.000168 & -- &-- &0.3209 &
0.0001679&$\sim 0$&K K \\
\hline
MP11&1 & $0,\pm 1$ & $0,\pm 1$ & $-0.00088$ & 0.28898 & -- & -- & 0.5024 &
0.1437 &$\sim 0$&K K\\
\hline
MP12&1 & 0 & $-1$ & $14.8319$ & 1 & -- & 0.0492 & 4.0738 & 0.2222 &$\sim
0$&K M\\
\hline
MP13&1 & 0 & $-1$ & $0.7335$ & 1 & -- & 1.807 & 0.9408 & 0.2222 &$\sim
0$&K M\\
\hline
MP14& 1 & $-1$ & 0 & 1 & 0.7181 & 0.8036 & -- &1 & 0.2044 &$\sim 0$&M K \\
\hline
MP15& 1 & $-1$ & 0 & 1 & 0.0417 & 0.9900 & -- & 1 & 0.0364 &$\sim
0$&M K\\
\hline
MP16& 1 & $-1$ & 0 & 1 & $-14.1607$ & 0.0851 & -- & 1 & 0.2916 &$\sim 0$&M K\\
\hline
MP17& 1 & $-1$ & $\pm 1$ & 1 & 0 & 1 & -- & 1 & 0&$\sim 0$&M K\\
\hline
MP18&1 & $\pm 1$ & $- 1$ & 0 & 1 & -- & 1 & 0.75 & 0.25&$\sim 0$& S$_\pm$ M\\
\hline
MP19&1 & $1$ & $-1$ & 1 & 1 & 0.2986 & 0.2958 & 1 & 0.25 &$\sim
0$&C M \\
\hline
MP20&1 & $-1$ & $-1$ & 1 & 1 & 1.329&0.7899 & 1 &  0.25&$\sim
0$&M M \\
\hline
MP21&1 & $-1$ & $-1$ & 1 & 1 & 1.112 & 1.957 & 1 &  0.25&$\sim
0$&M M \\
\hline
MP22&1 & $-1$ & $-1$ & 1 & 1 & 0.5677 & 0.9155 & 1 &  0.25&$\sim
0$& M M\\
\hline
\end{tabular}
}
\label{table_8}
\end{center}
\end{table}

Here we also focus on inflationary solutions.
In the original frame, ME1$_+$(MF5),  ME2$_+$(MF6),  ME3$_+$(MF7),
ME4$_+$(MF2),  ME5$_+$(MF3),  ME6$_+$(MF4) give an exponential expansion
for the external space. In the Einstein frame, we find either a
power-law inflation [ME1$_+$(MF5),  ME2$_+$(MF6),  ME3$_+$(MF7)]
or an exponential expansion [ME4$_+$(MF2),  ME5$_+$(MF3),  ME6$_+$(MF4)].
Just as the case of the heterotic strings,
we obtain strange solutions MP5 $\sim$ 7 and MP11, in which the external space
shrinks exponentially in the original frame, but it expands by a power law
in the Einstein frame.
In the past asymptotic solution MP12, we also find a power-law
inflation both in the original and Einstein frames.

\subsection{Concluding remarks}

Before we apply our solutions to cosmology,
we have to specify what kind of the universe we are looking for.
We wish to have an inflation in the early stage of the universe.
We also hope to find an accelerated expansion in the present stage,
if possible. Note that our cosmological model is higher-dimensional,
so that there are two kinds of frames that we can take to discuss
cosmologies, the original frame and the Einstein frame in four dimensions.
We must first determine in which frame we should consider the problem.
Notice that the flatness and horizon problems should be explained in our
four-dimensional spacetime for a successful inflationary scenario.
If the radius of the internal space does not change,
there is no difference between these frames.
On the other hand, the four-dimensional gravitational constant depends
on time in general unless we take the Einstein frame when the radius of the
internal space changes as in the present case, and this is not preferable
for a model of our universe. It thus appears more reasonable to consider
the problem in the Einstein frame.
Also the condition for the inflation in the Einstein frame is sufficient
for that in the original frame.
For these reasons, we require a successful inflation
in the Einstein frame. This may be regarded as a criterion for the
succesful inflation independent of the mechanism of fixing the size
of the internal space.

Next, we need at least 60 e-foldings of inflationary expansion.
This may give some constraint on the power exponent for a power-law inflation,
that is, the power exponent should be significantly larger than unity.
As we mentioned above, some solutions give an inflation in the Einstein
frame but the external space shrinks in the original higher dimensions.
Are such solutions suitable for a good cosmological model?
The answer is NO. Our four-dimensional universe makes sense only if
it is much larger than the internal space, so the external space should
expand faster than the internal space. Its expansion may not necessary
to be inflationary, but at least the external space must be expanding
in the whole space.

{}From these considerations and the above list of solutions,
we conclude that the M-theory (and type II string theories) can provide
successful inflationary solutions.
In the heterotic string theories with Gauss-Bonnet term, although we find
exponential expansions of the external space in the original frame,
those give non-inflationary power-law expansions in the Einstein frame.
There is a power-law inflation HP5 in the past asymptotic regime.
However, the power exponent is 3.3, which may be too small to solve
flatness and horizon problems, because we do not expect the expansion
in these solutions continues so long. We also have a super inflation
HE2$_+$ (HF4, HP4) in the Einstein frame~\cite{super_inflation}.
In this case, we have to clarify a mechanism to avoid the singularity
at $t_E=0$.

In the M-theory, we find seven candidates
(ME1$_+$, ME2$_+$, ME3$_+$, ME4$_+$, ME5$_+$, ME6$_+$ and MP12).
Among these, we can first exclude ME2$_+$ and ME3$_+$ because the internal
space expands almost at the same rate as the external space. As for the
solutions ME1$_+$ and MP12, we could also reject them because the power
exponents in the Einstein frame are 3.1 and 4.1, which may be too small.
However, this does not completely exclude the solutions because they may
give large e-foldings if the inflation lasts for long time. To check this,
we have to analyze how long such an inflationary period can continue.
In our previous paper~\cite{MO}, we showed that although the period may be
too short for the present value of $\tilde{\alpha}_4$ in the solution
ME1$_+$, if we change the coupling constant slightly,
we find a successful inflationary scenario with large extra dimensions.
Such change of coupling constant is possible because there is intrinsic
ambiguity in the terms of effective action involving Ricci tensors and
scalar curvature.

For the solutions ME4$_+$,ME5$_+$,ME6$_+$, we find an exponential expansion
of the external space
both in the original and the Einstein frames, and the internal space is static.
Hence those solutions may provide a successful inflationary scenario.

Which solution is preferable?
In order to answer this question, we have to analyze the dynamics of our
system. Then we should study the stability of those solutions
both perturbatively and non-perturbatively and find how large e-folding
we can get. This study is in progress and the results will be reported
in the forthcoming paper~\cite{AMO}.

\section*{Acknowledgments}

We would like to thank K. Akune, N. Deruelle, Y. Hyakutake, T. Shiromizu, T.
Torii, D. Wands, M. Yamaguchi and J. Yokoyama for useful  discussions. This
work was partially supported by the Grant-in-Aid for Scientific Research Fund
of the MEXT (Nos. 14540281 and 16540250) and by the Waseda University
Grant for Special Research Projects and  for The 21st Century
COE Program (Holistic Research and Education Center for Physics
Self-organization Systems) at Waseda University.

\appendix

\section{Equations of motion}

Taking variation of the actions, we find the basic equations~\p{eq1}
-- \p{eq3}, where each term is summarized here according to which action
it originates from. We use the following notation throughout this paper.
\bea
(p-m)_n&\equiv& (p-m)(p-m-1)(p-m-2)\cdots (p-n),
\nonumber \\
(q-m)_n&\equiv&(q-m)(q-m-1)(q-m-2)\cdots (q-n),
\nonumber \\
A_p&\equiv&\dot{u}_1^2+\sigma_p e^{2(u_0-u_1)}, \quad
A_q \equiv \dot{u}_2^2+\sigma_q e^{2(u_0-u_2)},\nn
X &\equiv& \ddot u_1 - \dot u_0 \dot u_1 +\dot u_1^2, \quad
Y \equiv \ddot u_2 - \dot u_0 \dot u_2 +\dot u_2^2.
\label{xy}
\ena
The Einstein equations are given by the following three equations:
\bea
&&F=0\\
&&F^{(p)}=0\\
&&F^{(q)}=0
\ena
where
\bea
F&=&{\partial S\over \partial u_0}=\sum_{n=1}^{4}F_n+F_S,\\
F^{(p)}&=&{1\over p}{\partial S\over \partial
u_1}=\sum_{n=1}^{4}F_n^{(p)}+F_S^{(p)},\\ F^{(q)}&=&{1\over q}{\partial S\over
\partial u_2}=\sum_{n=1}^{4}F_n^{(q)}+F_S^{(q)}
\ena
with
\bea
F_n^{(p)}&=&f_n^{(p)}+g_n^{(p)} X+h_n^{(p)}Y\\
F_n^{(q)}&=&f_n^{(q)}+g_n^{(q)} Y+h_n^{(q)}X\,.
\ena
$f_n^{(p)}, g_n^{(p)}, h_n^{(p)},f_n^{(q)}, g_n^{(q)}$, and $h_n^{(q)}$
are functionals of
$u_0,\dot{u}_1,\dot{u}_2,A_p$, and $A_q$, while $F_S, F_S^{(p)}$, and $F_S^{(q)}$
are functionals of $u_0,u_1,u_2,\dot{u}_0,\dot{u}_1,
\dot{u}_2,\ddot{u}_1,\ddot{u}_2,\dddot{u}_1,\dddot{u}_2,
X,Y,\dot{X},\dot{Y},\ddot{X}$, and $\ddot{Y}$.
The explicit forms of each term are listed here:

{\bf (1) EH action ($n=1$)}
\bea
\label{eh1}
F_1&=&\alpha_1 e^{-u_0}\left[p_1 A_p+q_1 A_q+2pq\dot{u}_1\dot{u}_2\right],
\\
f_1^{(p)}&=&\alpha_1 e^{-u_0}\left[
(p-1)_2A_p+q_1 A_q+2(p-1)q\dot{u}_1\dot{u}_2\right],
\\
f_1^{(q)}&=&\alpha_1 e^{-u_0}\left[p_1 A_p+
(q-1)_2A_q+2p(q-1)\dot{u}_1\dot{u}_2\right],
\\
g_1^{(p)}&=&2(p-1)\alpha_1 e^{-u_0},
\\
g_1^{(q)}&=&2(q-1)\alpha_1 e^{-u_0},
\\
h_1^{(p)}&=&2q\alpha_1 e^{-u_0},
\\
h_1^{(q)}&=&2p\alpha_1 e^{-u_0}.
\ena
{\bf (2) GB action ($n=2$)}
\bea
&&
\hs{-5} F_2 = \alpha_2 e^{-3u_0} [ p_3 A_p^2+2p_1q_1 A_pA_q+q_3 A_q^2
+4\dot{u}_1\dot{u}_2(p_2 q A_p+p q_2 A_q)
+ 4p_1q_1\dot{u}_1^2\dot{u}_2^2 ],\hs{5}
\\
&& f_2^{(p)}\; =\; \alpha_2 e^{-3u_0}\left[
(p-1)_4A_p^2+2(p-1)_2q_1 A_pA_q+q_3 A_q^2
\right.\nonumber \\
&& \hs{10}
\left.
+\; 4\dot{u}_1\dot{u}_2
\left((p-1)_3qA_p+(p-1)q_2
A_q\right)+4(p-1)_2q_1\dot{u}_1^2\dot{u}_2^2\right],
\\
&& f_2^{(q)}\; =\; \alpha_2 e^{-3u_0}\left[
(q-1)_4A_q^2+2(q-1)_2p_1 A_pA_q+p_3 A_p^2
\right.\nonumber \\
&& \hs{10}
\left.
+\; 4\dot{u}_1\dot{u}_2
\left((q-1)_3pA_q+(q-1)p_2
A_p\right)+4(q-1)_2p_1\dot{u}_1^2\dot{u}_2^2\right],
\\
&& g_2^{(p)}\; =\; 4(p-1)\alpha_2 e^{-3u_0}\left[
(p-2)_3A_p+q_1A_q+2(p-2)q\dot{u}_1\dot{u}_2\right],
\\
&& g_2^{(q)}\; =\; 4(q-1)\alpha_2 e^{-3u_0}\left[
(q-2)_3A_q+p_1A_p+2(q-2)p\dot{u}_1\dot{u}_2\right],
\\
&& h_2^{(p)}\; =\; 4q\alpha_2 e^{-3u_0}\left[
(p-1)_2A_p+(q-1)_2A_q+2(p-1)(q-1)\dot{u}_1\dot{u}_2\right],
\\
&& h_2^{(q)}\; =\; 4p\alpha_2 e^{-3u_0}\left[
(q-1)_2A_q+(p-1)_2A_p+2(p-1)(q-1)\dot{u}_1\dot{u}_2\right].
\label{gbl}
\ena
{\bf (3) Lovelock action ($n=3,4$)}
\bea
F_3&=& \alpha_3 e^{-5u_0}\left[
p_5 A_p^3+3p_3q_1 A_p^2A_q+3p_1q_3 A_pA_q^2+q_5 A_q^3
+6\dot{u}_1\dot{u}_2(p_4 q A_p^2
\right.
\nonumber \\
&&
\left.
\hs{15}
+2p_2q_2A_pA_q+p q_4 A_q^2)
+12\dot{u}_1^2\dot{u}_2^2(
p_3q_1A_p+p_1q_3A_q)+8p_2q_2\dot{u}_1^3\dot{u}_2^3 \right],
\\
f_3^{(p)}&=&\alpha_3 e^{-5u_0}\left[
(p-1)_6 A_p^3+3(p-1)_4q_1 A_p^2A_q+3(p-1)_2q_3A_pA_q^2+q_5 A_q^3
\right.\nonumber \\
&&
\left.
+6\dot{u}_1\dot{u}_2
\left((p-1)_5qA_p^2+2(p-1)_3q_2A_pA_q+(p-1)q_4 A_q^2\right)
\right.
\nonumber \\
&&\left.
+12\dot{u}_1^2\dot{u}_2^2\left((p-1)_4q_1A_p+(p-1)_2q_3A_q\right)
+8(p-1)_3q_2\dot{u}_1^3\dot{u}_2^3 \right],
\\
f_3^{(q)}&=&\alpha_3 e^{-5u_0}\left[
(q-1)_6 A_q^3+3(q-1)_4p_1 A_q^2A_p+3(q-1)_2p_3A_qA_p^2+p_5 A_p^3
\right.\nonumber \\
&&
\left.
+6\dot{u}_1\dot{u}_2
\left((q-1)_5pA_q^2+2(q-1)_3p_2A_qA_p+(q-1)p_4
A_p^2\right)
\right.
\nonumber \\
&&\left.
+12\dot{u}_1^2\dot{u}_2^2\left((q-1)_4p_1A_q+(q-1)_2p_3A_p\right)
+8(q-1)_3p_2\dot{u}_1^3\dot{u}_2^3 \right],
\\
g_3^{(p)}&=&6(p-1)\alpha_3 e^{-5u_0}\left[
(p-2)_5A_p^2+2(p-2)_3q_1A_pA_q+q_3A_q^2
\right.\nonumber\\
&&
\left.
+ 4\dot{u}_1\dot{u}_2\left(
(p-2)_4qA_p+(p-2) q_2A_q\right)
+4(p-2)_3q_1\dot{u}_1^2\dot{u}_2^2\right],
\\
g_3^{(q)}&=&6(q-1)\alpha_3 e^{-5u_0}\left[
(q-2)_5A_q^2+2(q-2)_3p_1A_pA_q+p_3A_p^2
\right.\nonumber\\
&&
\left.
+ 4\dot{u}_1\dot{u}_2\left(
(q-2)_4pA_q+(q-2) p_2A_p\right)
+4(q-2)_3p_1\dot{u}_1^2\dot{u}_2^2\right],
\\
h_3^{(p)}&=&6q\alpha_3 e^{-5u_0}\left[
(p-1)_4A_p^2+2(p-1)_2(q-1)_2A_pA_q
+(q-1)_4A_q^2
\right.
\nonumber\\
&&\left.
+4\dot{u}_1\dot{u}_2\left(
(p-1)_3(q-1)A_p+(p-1)(q-1)_3A_q\right)
+4(p-1)_2(q-1)_2\dot{u}_1^2\dot{u}_2^2\right],
\\
h_3^{(q)}&=&6p\alpha_3 e^{-5u_0}\left[
(q-1)_4A_q^2+2(q-1)_2(p-1)_2A_pA_q +(p-1)_4A_p^2
\right.
\nonumber\\
&&\left.
+4\dot{u}_1\dot{u}_2\left(
(q-1)_3(p-1)A_q+(q-1)(p-1)_3A_p\right)
+4(p-1)_2(q-1)_2\dot{u}_1^2\dot{u}_2^2\right],
\\
F_4&=&
\alpha_4 e^{-7u_0} \left[
p_7 A_p^4+4p_5q_1 A_p^3A_q+6p_3q_3A_p^2A_q^2
+4p_1q_5 A_pA_q^3+q_7 A_q^4
\right.
\nonumber \\
&&
+8\dot{u}_1\dot{u}_2
(p_6 q A_p^3+3p_4q_2A_p^2A_q+3p_2q_4A_pA_q^2 +p q_6 A_q^3)
\nonumber
 \\
&&\left.
+24\dot{u}_1^2\dot{u}_2^2(p_5q_1A_p^2+2p_3q_3A_pA_q+p_1q_5A_q^2)
+32\dot{u}_1^3\dot{u}_2^3(p_4q_2A_p+p_2q_4A_q)
+16p_3q_3\dot{u}_1^4\dot{u}_2^4 \right],
\nonumber \\
~~
\\
f_4^{(p)}&=&\alpha_4 e^{-7u_0}\left[ (p-1)_8 A_p^4+4(p-1)_6q_1
A_p^3A_q+6(p-1)_4q_3A_p^2A_q^2 +4(p-1)_2q_5A_pA_q^3+q_7 A_q^4
\right.\nonumber \\
&&
\left.
+8\dot{u}_1\dot{u}_2 \left((p-1)_7qA_p^3+3(p-1)_5q_2A_p^2A_q
+3(p-1)_3q_4A_pA_q^2+(p-1)q_6 A_q^3\right)
\right.
\nonumber \\
&&\left.
+24\dot{u}_1^2\dot{u}_2^2\left((p-1)_6q_1A_p^2+2(p-1)_4q_3A_pA_q
+(p-1)_2q_5A_q^2\right)
\right.
\nonumber \\
&&
\left.
+32\dot{u}_1^3\dot{u}_2^3\left( (p-1)_5q_2A_p+(p-1)_3q_4 A_q\right)
+16(p-1)_4q_3\dot{u}_1^4\dot{u}_2^4\right],
\\
f_4^{(q)}&=&\alpha_4 e^{-7u_0}\left[ (q-1)_8 A_q^4+4(q-1)_6p_1
A_q^3A_p+6(q-1)_4p_3A_p^2A_q^2 +4(q-1)_2p_5A_qA_p^3+p_7 A_p^4
\right.\nonumber \\
&&
\left.
+8\dot{u}_1\dot{u}_2 \left((q-1)_7pA_q^3+3(q-1)_5p_2A_q^2A_p
+3(q-1)_3p_4A_qA_p^2+(q-1)p_6 A_p^3\right)
\right.
\nonumber \\
&&\left.
+24\dot{u}_1^2\dot{u}_2^2\left((q-1)_6p_1A_q^2+2(q-1)_4p_3A_pA_q
+(q-1)_2p_5A_p^2\right)
\right.
\nonumber \\
&&
\left.
+32\dot{u}_1^3\dot{u}_2^3\left( (q-1)_5p_2A_q+(q-1)_3p_4 A_p\right)
+16(q-1)_4p_3\dot{u}_1^4\dot{u}_2^4 \right],
\\
g_4^{(p)}&=&8(p-1)\alpha_4 e^{-7u_0}\left[
(p-2)_7A_p^3+3(p-2)_5q_1A_p^2A_q+3(p-2)_3q_3A_pA_q^2+q_5A_q^3
\right.\nonumber\\
&&
\left.
+6\dot{u}_1\dot{u}_2\left( (p-2)_6qA_p^2+2(p-2)_4q_2A_pA_q
+(p-2)q_4A_q^2\right)\right.
\nonumber\\
&&
\left.
+12\dot{u}_1^2\dot{u}_2^2\left( (p-2)_5q_1A_p+(p-2)_3q_3A_q\right)
+8(p-2)_4q_2\dot{u}_1^3\dot{u}_2^3\right],
\\
g_4^{(q)}&=&8(q-1)\alpha_4 e^{-7u_0}\left[
(q-2)_7A_q^3+3(q-2)_5p_1A_q^2A_p+3(q-2)_3p_3A_qA_p^2+p_5A_p^3
\right.\nonumber\\
&&
\left.
+6\dot{u}_1\dot{u}_2\left( (q-2)_6pA_q^2+2(q-2)_4p_2A_pA_q
+(q-2)p_4A_p^2\right)\right.
\nonumber\\
&&
\left.
+12\dot{u}_1^2\dot{u}_2^2\left( (q-2)_5p_1A_q+(q-2)_3p_3A_p\right)
+8(q-2)_4p_2\dot{u}_1^3\dot{u}_2^3\right],
\\
h_4^{(p)}&=&8q\alpha_4 e^{-7u_0}\left[
(p-1)_6A_p^3+3(p-1)_4(q-1)_2A_p^2A_q+3(p-1)_2(q-1)_4A_pA_q^2
+(q-1)_6A_q^3
\right.
\nonumber\\
&&\left.
+6\dot{u}_1\dot{u}_2\left( (p-1)_5(q-1)A_p^2+2(p-1)_3(q-1)_3A_pA_q
+(p-1)(q-1)_5A_q^2\right)
\right.
\nonumber \\
&&
\left.
+12\dot{u}_1^2\dot{u}_2^2\left( (p-1)_4(q-1)_2A_p+(p-1)_2(q-1)_4A_q\right)
+8(p-1)_3(q-1)_3\dot{u}_1^3\dot{u}_2^3\right],
\\
h_4^{(q)}&=&8p\alpha_4 e^{-7u_0}\left[
(q-1)_6A_q^3+3(q-1)_4(p-1)_2A_q^2A_p+3(q-1)_2(p-1)_4A_qA_p^2 +(p-1)_6A_p^3
\right.
\nonumber\\
&&\left.
+6\dot{u}_1\dot{u}_2\left( (q-1)_5(p-1)A_q^2+2(p-1)_3(q-1)_3A_pA_q
+(q-1)(p-1)_5A_p^2\right)
\right.
\nonumber \\
&&
\left.
+12\dot{u}_1^2\dot{u}_2^2\left( (q-1)_4(p-1)_2A_q+(q-1)_2(p-1)_4A_p\right)
+8(p-1)_3(q-1)_3\dot{u}_1^3\dot{u}_2^3\right].
\ena
{\bf (4) $S_S$ action}
\bea
&&
F_S=\c e^{-pu_1-qu_2} \left[-7L_4+2\sigma_p e^{2(u_0-u_1)}{\partial
L_4\over \partial A_p}+2\sigma_q e^{2(u_0-u_2)}{\partial L_4\over
\partial A_q}+{d\over dt}\left(\dot{u}_1{\partial L_4\over
\partial X}+\dot{u}_2{\partial L_4\over \partial Y}\right)\right]\,,
\nn
&&
\\
&&
pF_S^{(p)}=\c e^{-pu_1-qu_2} \left[
pL_4-2\sigma_p e^{2(u_0-u_1)}{\partial L_4\over \partial A_p}
+{d\over dt}\left((\dot{u}_0-2\dot{u}_1){\partial L_4\over \partial
X}-2\dot{u}_1{\partial L_4\over \partial A_p}-{\partial L_4\over \partial
\dot{u}_1}\right) \right. \nn
&& \hs{50}\left. +\; {d^2\over dt^2}\left({\partial L_4\over
 \partial X}\right) \right]\,,
\\
&&
qF_S^{(q)}=\c e^{-pu_1-qu_2} \left[
qL_4-2\sigma_q e^{2(u_0-u_2)}{\partial L_4\over \partial A_q}
+{d\over dt}\left((\dot{u}_0-2\dot{u}_2){\partial L_4\over \partial
Y}-2\dot{u}_2{\partial L_4\over
\partial A_q}-{\partial L_4\over \partial \dot{u}_2}\right)\right.\nn
&& \hs{50} \left. + {d^2\over dt^2}\left({\partial L_4\over\partial Y}\right)
\right]
\,,
\ena
where
\bea
&&
L_4=e^{-7u_0+pu_1+qu_2}\left[p_1 X^2(X+2A_p)^2+q_1 Y^2(Y+2A_q)^2
+2pq(XY+(X+Y)\dot{u}_1\dot{u}_2)^2
\right.
\nn
&&
~~~~~
+3p_2 A_p^4+3q_2 A_q^4
\left.
+p_1 q\dot{u}_1^2\dot{u}_2^2(\dot{u}_1\dot{u}_2+2A_p)^2
+pq_1 \dot{u}_1^2\dot{u}_2^2(\dot{u}_1\dot{u}_2+2A_q)^2
\right],
\\
&&
{\partial L_4\over\partial X}=
4pe^{-7u_0+pu_1+qu_2}\left[
(p-1)X(X+A_p)(X+2A_p)+q(Y+\dot{u}_1\dot{u}_2)(XY+(X+Y)\dot{u}_1\dot{u}_2)
\right],
\nn
&&
\\
&&
{\partial L_4\over \partial Y}=
4qe^{-7u_0+pu_1+qu_2}\left[
(q-1)Y(Y+A_q)(Y+2A_q)+p(X+\dot{u}_1\dot{u}_2)(XY+(X+Y)\dot{u}_1\dot{u}_2)
\right],
\nn
&&
\\
&&
{\partial L_4\over \partial A_p}=
4p_1 e^{-7u_0+pu_1+qu_2}\left[
X^2(X+2A_p)+3(p-2)A_p^3+q\dot{u}_1^2\dot{u}_2^2(\dot{u}_1\dot{u}_2+2A_p)
\right],
\\
&&
{\partial L_4\over \partial A_q}=
4q_1 e^{-7u_0+pu_1+qu_2}\left[
Y^2(Y+2A_q)+3(q-2)A_q^3+p\dot{u}_1^2\dot{u}_2^2(\dot{u}_1\dot{u}_2+2A_q)
\right],
\\
&&
{\partial L_4\over \partial\dot{u}_1}=
4pq~e^{-7u_0+pu_1+qu_2}~\dot{u}_2\left[
(X+Y)(XY+(X+Y)\dot{u}_1\dot{u}_2)
\right.
\nn
&&
\left.
~~~~~+(p-1)\dot{u}_1\dot{u}_2
(\dot{u}_1\dot{u}_2+A_p)(\dot{u}_1\dot{u}_2+2A_p)
+(q-1)\dot{u}_1\dot{u}_2
(\dot{u}_1\dot{u}_2+A_q)(\dot{u}_1\dot{u}_2+2A_q)
\right],
\\
&&
{\partial L_4\over\partial \dot{u}_2}=
4pq~e^{-7u_0+pu_1+qu_2}~\dot{u}_1\left[
(X+Y)(XY+(X+Y)\dot{u}_1\dot{u}_2)
\right.
\nn
&&
\left.
~~~~~+(p-1)\dot{u}_1\dot{u}_2
(\dot{u}_1\dot{u}_2+A_p)(\dot{u}_1\dot{u}_2+2A_p)
+(q-1)\dot{u}_1\dot{u}_2
(\dot{u}_1\dot{u}_2+A_q)(\dot{u}_1\dot{u}_2+2A_q)
\right].
\ena

\section{Inputting our ansatz into solutions}

In order to find solutions, we assume
\bea
u_0=\epsilon t \,, u_1=\mu t  + \ln a_0\,, u_2=\nu t  + \ln b_0 \,.
\ena
Inserting this form into the above equations (Eqs.~\p{eh1} -- \p{gbl})
and setting
\bea
&& A_p=\mu^2+\tilde{\sigma}_p e^{2(\epsilon-\mu)t}, ~~~
A_q=\nu^2+\tilde{\sigma}_q e^{2(\epsilon-\nu)t}, ~~~
\tilde\s_p =\frac{\s_p}{a_0^2}, ~~~
\tilde\s_q =\frac{\s_q}{b_0^2},
\label{A_pA_q}
\\
&&
X=\mu(\mu-\epsilon), ~~~
Y=\nu(\nu-\epsilon).
\label{XY}
\ena

we obtain
\bea
F_1&=&\alpha_1 e^{-\epsilon t}\left[p_1A_p+q_1A_q+2pq \mu\nu\right],
\nn
f_1^{(p)}&=&\alpha_1 e^{-\epsilon t}\left[(p-1)_2A_p+q_1A_q+2(p-1)q
\mu\nu\right],
\nn
f_1^{(q)}&=&\alpha_1 e^{-\epsilon t}\left[p_1A_p+(q-1)_2A_q+2p(q-1)
\mu\nu\right],
\nn
g_1^{(p)}&=&2(p-1)\alpha_1 e^{-\epsilon t}
, ~~~g_1^{(q)}~=~2(q-1)\alpha_1 e^{-\epsilon t},
\nn
h_1^{(p)}&=&2q\alpha_1 e^{-\epsilon t}
, ~~~~~~~~~~~
h_1^{(q)}~=~2p\alpha_1 e^{-\epsilon t},
\label{gbeq1}
\\
F_2&=&\alpha_2 e^{-3\epsilon t}\left[
p_3A_p^2+2p_1q_1A_pA_q+q_3A_q^2+4\mu\nu\left(
p_2qA_p+pq_2A_q\right)+4p_1q_1\mu^2\nu^2\right],
\nn
f_2^{(p)}&=&\alpha_2 e^{-3\epsilon t}\left[
(p-1)_4A_p^2+2(p-1)_2q_1A_pA_q+q_3A_q^2
\right.
\nn
&&
\left.
+4\mu\nu\left(
(p-1)_3qA_p+(p-1)q_2A_q\right)+4(p-1)_2q_1\mu^2\nu^2 \right],
\nn
f_2^{(q)}&=&\alpha_2 e^{-3\epsilon t}\left[
(q-1)_4A_q^2+2(q-1)_2p_1A_pA_q+p_3A_p^2
\right.
\nn
&&
\left.
+4\mu\nu\left(
(q-1)_3pA_q+(q-1)p_2A_p\right)+4(q-1)_2p_1\mu^2\nu^2 \right],
\nn
g_2^{(p)}&=&4(p-1)\alpha_2 e^{-3\epsilon t}\left[
(p-2)_3A_p+q_1A_q+2(p-2)q\mu\nu \right],
\nn
g_2^{(q)}&=&4(q-1)\alpha_2 e^{-3\epsilon t}\left[
(q-2)_3A_q+p_1A_p+2(q-2)p\mu\nu \right],
\nn
h_2^{(p)}&=&4q\alpha_2 e^{-3\epsilon t}\left[
(p-1)_2A_p+(q-1)_2A_q+2(p-1)(q-1)\mu\nu \right],
\nn
h_2^{(q)}&=&4p\alpha_2 e^{-3\epsilon t}\left[
(p-1)_2A_p+(q-1)_2A_q+2(p-1)(q-1)\mu\nu
\right]\,,
\label{gbeq2}
\ena
\bea
F_4&=&\alpha_4 e^{-7\epsilon t}
\left[ p_7 A_p^4+4p_5q_1 A_p^3A_q+6p_3q_3A_p^2A_q^2
+4p_1q_5 A_pA_q^3+q_7 A_q^4 \right. \nn
&&
+8\mu\nu (p_6 q A_p^3+3p_4q_2A_p^2A_q+3p_2q_4A_pA_q^2 +p q_6 A_q^3)
+24\mu^2\nu^2(p_5q_1A_p^2\nn
&&\left.
+2p_3q_3A_pA_q+p_1q_5A_q^2)
+32\mu^3\nu^3(p_4q_2A_p+p_2q_4A_q)
+16p_3q_3\mu^4\nu^4 \right],
\label{M:F_4}
\\
f_4^{(p)}&=&\alpha_4 e^{-7\epsilon t}\Big[
(p-1)_8 A_p^4+4(p-1)_6q_1 A_p^3A_q+6(p-1)_4q_3A_p^2A_q^2
+4(p-1)_2q_5A_pA_q^3 \nn
&&
+q_7 A_q^4 +8\mu\nu\left\{(p-1)_7qA_p^3+3(p-1)_5q_2A_p^2A_q
+3(p-1)_3q_4A_pA_q^2 \right. \nn
&& \left. +(p-1)q_6 A_q^3\right\}
+24\mu^2\nu^2\left\{(p-1)_6q_1A_p^2+2(p-1)_4q_3A_pA_q
+(p-1)_2q_5A_q^2\right\} \nn
&& +32\mu^3\nu^3\left\{(p-1)_5q_2A_p+(p-1)_3q_4 A_q\right\}
+16(p-1)_4q_3\mu^4\nu^4 \Big],
\label{M:f_4p}
\\
g_4^{(p)}&=&8(p-1)\alpha_4 e^{-7\epsilon t}\left[
(p-2)_7A_p^3+3(p-2)_5q_1A_p^2A_q+3(p-2)_3q_3A_pA_q^2+q_5A_q^3
\right.\nn
&&
\left.
+6\mu\nu\left\{ (p-2)_6qA_p^2+2(p-2)_4q_2A_pA_q
+(p-2)q_4A_q^2\right\}\right. \nn
&& \left. +12\mu^2\nu^2\left\{ (p-2)_5q_1A_p+(p-2)_3q_3A_q\right\}
+8(p-2)_4q_2\mu^3\nu^3\right],
\label{M:g_4p}
\\
h_4^{(p)}&=&8q\alpha_4 e^{-7\epsilon t}\left[
(p-1)_6A_p^3+3(p-1)_4(q-1)_2A_p^2A_q+3(p-1)_2(q-1)_4A_pA_q^2 \right. \nn
&& +(q-1)_6A_q^3 + 6\mu\nu\left\{ (p-1)_5(q-1)A_p^2+2(p-1)_3(q-1)_3A_pA_q
\right.\nn
&& \left. +(p-1)(q-1)_5A_q^2\right\} +12\mu^2\nu^2\left\{
(p-1)_4(q-1)_2A_p+(p-1)_2(q-1)_4A_q\right\}\nn
&& +8(p-1)_3(q-1)_3\mu^3\nu^3\Big],
\label{M:h_4p}
\\
f_4^{(q)}&=&\alpha_4 e^{-7\epsilon t}\Big[ (q-1)_8 A_q^4+4(q-1)_6p_1
A_q^3A_p+6(q-1)_4p_3A_p^2A_q^2 +4(q-1)_2p_5A_qA_p^3 \nn
&&
+p_7 A_p^4+8\mu\nu
\left\{(q-1)_7pA_q^3+3(q-1)_5p_2A_q^2A_p+3(q-1)_3p_4A_qA_p^2 \right.\nn
&& \left. +(q-1)p_6 A_p^3\right\}+24\mu^2\nu^2\left\{(q-1)_6p_1A_q^2
+2(q-1)_4p_3A_pA_q +(q-1)_2p_5A_p^2\right\} \nn
&&
\left.
+32\mu^3\nu^3\left\{
(q-1)_5p_2A_q+(q-1)_3p_4 A_p\right\}
+16(q-1)_4p_3\mu^4\nu^4 \right],
\label{M:f_4q}
\\
g_4^{(q)}&=&8(q-1)\alpha_4 e^{-7\epsilon t}\left[
(q-2)_7A_q^3+3(q-2)_5p_1A_q^2A_p+3(q-2)_3p_3A_qA_p^2+p_5A_p^3
\right.\nonumber\\
&&
+6\mu\nu\left\{ (q-2)_6pA_q^2+2(q-2)_4p_2A_pA_q
+(q-2)p_4A_p^2\right\} \nn
&&
\left.
+12\mu^2\nu^2\left\{ (q-2)_5p_1A_q+(q-2)_3p_3A_p\right\}
+8(q-2)_4p_2\mu^3\nu^3\right],
\label{M:g_4q}
\\
h_4^{(q)}&=&8p\alpha_4 e^{-7\epsilon t}\Big[
(q-1)_6A_q^3+3(q-1)_4(p-1)_2A_q^2A_p+3(q-1)_2(p-1)_4A_qA_p^2 \nn
&& +(p-1)_6A_p^3 +6\mu\nu\left\{ (q-1)_5(p-1)A_q^2+2(p-1)_3(q-1)_3A_pA_q
\right. \nn
&& \left. +(q-1)(p-1)_5A_p^2\right\}
+12\mu^2\nu^2\left\{(q-1)_4(p-1)_2A_q+(q-1)_2(p-1)_4A_p\right\} \nn
&& +8(p-1)_3(q-1)_3\mu^3\nu^3\Big],
\label{M:h_4q}
\\
F_S&=&\gamma e^{-7\epsilon t}\left[
-7\tilde{L}_4+\left(-7\epsilon +p\mu+q\nu\right)\left(p\mu M_X +q\nu
M_Y\right)
\right.
\nn
&&
+2\tilde{\sigma}_p
e^{2(\epsilon-\mu)t} p \left\{N_p+\mu(\epsilon-\mu) P_{pX}\right\}
+2\tilde{\sigma}_q e^{2(\epsilon-\nu)t} q \left\{N_q+\nu(\epsilon-\nu)
P_{qY}\right\}\Big],
\label{M:F_S}
\\
F_S^{(p)}&=&\gamma e^{-7\epsilon t} \Big[ \tilde{L}_4
+(-7\epsilon+p\mu+q\nu)\left\{(\epsilon-2\mu)M_X-2\mu N_p-q \nu U \right.
\nn
&&
\left.+4(\epsilon-\mu)\tilde{\sigma}_p e^{2(\epsilon-\mu)t}P_{pX}\right\}
+(-7\epsilon+p\mu+q\nu)^2M_X+2\tilde{\sigma}_p
e^{2(\epsilon-\mu)t}\Big\{-N_p \nn
&&
+(\epsilon-\mu)\left[(\epsilon-2\mu)P_{pX}-2\mu Q_{pp}-(p-1)q \nu  V_p
\right]\Big\} -2(\epsilon-\nu)\tilde{\sigma}_q e^{2(\epsilon-\nu)t} q_1
\nu V_q  \nn && +4(\epsilon-\mu)^2\tilde{\sigma}_p e^{2(\epsilon-\mu)t}
P_{pX} +4(\epsilon-\mu)^2\tilde{\sigma}_p^2 e^{4(\epsilon-\mu)t}
R_{pX}\Big],
\label{M:F_Sp}
\\
F_S^{(q)}&=&\gamma e^{-7\epsilon t}\Big[ \tilde{L}_4
+(-7\epsilon+p\mu+q\nu)\left\{(\epsilon-2\nu)M_Y-2\nu N_q-p\mu U \right. \nn
&&
\left. +4(\epsilon-\nu)\tilde{\sigma}_q e^{2(\epsilon-\nu)t}P_{qY}\right\}
+(-7\epsilon+p\mu+q\nu)^2M_Y+2\tilde{\sigma}_q
e^{2(\epsilon-\nu)t} \Big\{-N_q \nn
&& +(\epsilon-\nu)\left[(\epsilon-2\nu)P_{qY}-2\nu Q_{qq}- p(q-1) \mu
V_q \right]
\Big\}
-2(\epsilon-\mu)\tilde{\sigma}_p e^{2(\epsilon-\mu)t} p_1 \mu  V_p  \nn
&& +4(\epsilon-\nu)^2\tilde{\sigma}_q e^{2(\epsilon-\nu)t}
P_{qY}+4(\epsilon-\nu)^2\tilde{\sigma}_q^2 e^{4(\epsilon-\nu)t} R_{qY}
\Big]\,,
\label{M:F_Sq}
\ena
where
\bea
\tilde{L}_4&=& p_1 X^2(X+2A_p)^2
+q_1Y^2(Y+2A_q)^2+2pq(XY+(X+Y)\mu\nu)^2 \nn
&&
+3p_2A_p^4+3q_2A_q^4 + p_1q\mu^2\nu^2(\mu\nu+2A_p)^2
+pq_1\mu^2\nu^2(\mu\nu+2A_q)^2
,
\\
M_X&=&4\left[
(p-1)X(X+A_p)(X+2A_p)+q(Y+\mu\nu)(XY+(X+Y)\mu\nu) \right],
\\
M_Y&=&4\left[
(q-1)Y(Y+A_q)(Y+2A_q)+p(X+\mu\nu)(XY+(X+Y)\mu\nu) \right],
\\
N_p&=&4(p-1)\left[
X^2(X+2A_p)+3(p-2)A_p^3+q\mu^2\nu^2(\mu\nu+2A_p) \right],
\\
N_q&=&4(q-1)\left[
Y^2(Y+2A_q)+3(q-2)A_q^3+p\mu^2\nu^2(\mu\nu+2A_q) \right],
\\
P_{pX}&=&4(p-1)X(3X+4A_p),
\\
P_{qY}&=&4(q-1)Y(3Y+4A_q),
\\
Q_{pp}&=&4(p-1)\left[
2X^2+9(p-2)A_p^2+2q\mu^2\nu^2 \right],
\\
Q_{qq}&=&4(q-1)\left[
2Y^2+9(q-2)A_q^2+2p\mu^2\nu^2 \right],
\\
R_{pX} &=&16(p-1)X,
\\
R_{qY} &=&16(q-1)Y,
\\
U &=& 4\left[ (X+Y)(XY+(X+Y)\mu\nu)
\right.
\nn
&&
\left.
+(p-1)\mu\nu(\mu\nu+A_p)(\mu\nu+2A_p)
+(q-1)\mu\nu(\mu\nu+A_q)(\mu\nu+2A_q)\right],
\\
V_p &=& 4 \mu\nu(3\mu\nu+4A_p)\,, \\
V_q &=& 4 \mu\nu(3\mu\nu+4A_q)\,.
\ena

\section{Generalized de Sitter solutions with flat spaces}

If we assume  $\epsilon=0$, $\sigma_p=0$, and $\sigma_q=0$,
we find two independent basic equations without constraint.
Setting  $X=\mu^2=A_p$ and $Y=\nu^2=A_q$, we obtain the following
terms:
\bea
F_1&=&\alpha_1\left[p_1\mu^2+q_1 \nu^2+2pq\mu\nu \right],
\\
F_1^{(p)} &=& F_1-q\nu H_1, \\
F_1^{(q)} &=& F_1+p\mu H_1, \\
F_2 &=& \alpha_2\left[ p_3 \mu^4+ 4\mu\nu ( p_2 q \mu^2+p q_2 \nu^2 )
+6 p_1 q_1 \mu^2 \nu^2+q_3 \nu^4 \right],
\\
F_2^{(p)} &=& F_2-q\nu H_2, \\
F_2^{(q)} &=& F_2+p\mu H_2, \\
F_4&=&\alpha_4 \big[p_7 \mu^8+8p_6 q\mu^7\nu+28p_5q_1 \mu^6\nu^2+56 p_4
q_2\mu^5\nu^3+70p_3q_3\mu^4\nu^4 \nn
&&
~~~~~+56 p_2 q_4 \mu^3\nu^5
+28p_1q_5 \mu^2\nu^6+8pq_6 \mu\nu^7+q_7 \nu^8\big],
\label{MdS:F_4}
\\
F_4^{(p)}&=&F_4-q\nu H_4,
\label{MdS:F_4p} \\
F_4^{(q)}&=&F_4+p\mu H_4,
\label{MdS:F_4q} \\
F_S &=& \gamma \Big[
-7\tilde{L}_4+\left(p\mu+q\nu\right)\left(p\mu M_X+q\nu M_Y\right) \Big],
\label{MdS:F_S}
\\
F_S^{(p)}&=&F_S-\gamma q\nu H_S
\label{MdS:F_Sp}
\\
F_S^{(q)}&=&F_S+\gamma p\mu H_S \,,
\label{MdS:F_Sq}
\ena
where
\bea
H_1 &=& 2\alpha_1(\mu-\nu),
\\
H_2 &=& 4\alpha_2 (\mu-\nu) \left[ (p-1)_2 \mu^2 + 2(p-1)(q-1)\mu\nu
+(q-1)_2 \nu^2 \right],
\\
H_4 &=& 8\alpha_4 (\mu-\nu) \left[ (p-1)_6\mu^6 + 6(p-1)_5(q-1) \mu^5\nu
+15(p-1)_4(q-1)_2 \mu^4\nu^2 \right. \nn
&&
+ 20(p-1)_3(q-1)_3 \mu^3\nu^3 + 15 (p-1)_2(q-1)_4 \mu^2\nu^4 \nn
&&
\left. + 6(p-1)(q-1)_5 \mu\nu^5 + (q-1)_6 \nu^6 \right],
\\
\tilde{L}_4 &=& 3(p+1)_1 \mu^8 +3(q+1)_1\nu^8
+2pq(\mu^2+\nu^2+\mu\nu)^2\mu^2\nu^2 \nn
&&
+p_1q\mu^4\nu^2(2\mu+\nu)^2 +pq_1\mu^2\nu^4(\mu+2\nu)^2,
\\
M_X&=&4\left[6(p-1)\mu^6+q\mu\nu^2(\mu+\nu)(\mu^2+\nu^2+\mu\nu)\right],
\\
M_Y&=&4\left[6(q-1)\nu^6+p\mu^2\nu(\mu+\nu)(\mu^2+\nu^2+\mu\nu)\right],
\\
H_S &=& 4\gamma (\mu-\nu) \Big[ 6 (p-1)\mu^6
- (p-1)(p+6q-6) \mu^5 \nu -(2p^2-7p-3q+6)\mu^4 \nu^2 \nn
&& - \{p^2-(4q+3)p+q^2-3q+6\} \mu^3\nu^3 -(2q^2-7q-3p+6)\mu^2 \nu^4\nn
&& - (q-1)(q+6p-6) \mu\nu^5 +6(q-1)\nu^6 \Big].
\ena
Since the second and third equations are given in the form
\bea
F^{(p)}=F-q\nu H, ~~~{\rm and }~~~F^{(p)}=F+p\mu H
\ena
those are are equivalent if $\mu\neq 0$ and $\nu\neq 0$.
Then we can take the following two algebraic equations  as our basic equations.
\bea
&&\sum_{n=1}^4 F_n+F_S=0\,,
\\
&&\sum_{n=1}^4 H_n+H_S=0\,.
\ena

\section{Solutions in type II string}
Here we summarize the case of Type II string.
Exact solutions are in Table~\ref{table_9},
future asymptotic solutions in Table~\ref{table_10}
and past asymptotic solutions in Table~\ref{table_11}.

Here we have the similar results to the case of the M-theory.
Let us focus on inflationary solutions.
In the original frame, IIE1$_+$(IIF4),  IIE2$_+$(IIF5),  IIE3$_+$(IIF6),
IIE4$_+$(IIF2),  IIE5$_+$(IIF3)    give an exponential expansion
for the external space. In the Einstein frame, we find either a
power-law inflation [IIE1$_+$(IIF4),  IIE2$_+$(IIF5),  IIE3$_+$(IIF6)]
or an exponential expansion [IIE4$_+$(IIF2),  IIE5$_+$(IIF3)].
Just as the case of the heterotic strings,
we obtain strange solutions IIP4$\sim$6, in which the external space
shrinks exponentially in the original frame, but it expands by a  power law
in the Einstein frame.
In some past asymptotic solutions [IIP7 and IIP11], we also find a power
law inflation both in the original and Einstein frames.

\begin{table}[H]
\caption{Type II superstring: exact solutions.
 K, S$_\pm$, S$_0$, and M mean
a kinetic dominance, a static space with positive (or negative) curvature,
a flat static space, and a Milne type space, respectively.}
\begin{center}
{\footnotesize
\begin{tabular}{|c||c||c|c||c|c|c|c||c|c||c|}
\hline
&$\epsilon$&$\sigma_p$&$\sigma_q$&$\mu$&$\nu$&$a_0$&$b_0$&
$\lambda$&$\phi_1$& type \\
\hline
\hline
IIE1$_\pm$&0 & 0 & 0 & $\pm 0.7999$ & $\pm 0.1299$ & -- & -- & 3.053 &
$\pm 0.0433 $&K K\\
\hline
IIE2$_\pm$&0 & 0 & 0 & $\pm 0.5075$ & $\pm 0.5075$ & -- & -- &1.333&$\pm
0.1692$&K K\\
\hline
IIE3$_\pm$&0 & 0 & 0 & $\pm 0.4962$ & $\pm 0.5131$ & -- & -- &1.322 & $\pm
0.1710$&K K
\\
\hline
IIE4$_\pm$&0 & 0 & 1 & $\pm 0.7655$ & 0 & -- & 1.480 & $e^{\mu t_E}$ & 0 &K
S$_+$
\\
\hline
IIE5$_\pm$&0 & 0 & $-1$ & $\pm 0.6200$ & 0 & -- &2.763 &  $e^{\mu t_E}$ & 0 &K
S$_-$
\\
\hline
IIE6$_\pm$&0 & 1 & 0 & 0 & $\pm 0.6201$ & 1.078 & -- & 1 & $\pm 0.2067$ &
S$_+$ K\\
\hline
\hline
IIE7&1 & 0 & $-1$ & 0 & 1 &-- & 1 & 0.75 & 0.25  &
S$_0$ M\\
\hline
IIE8&1 & $-1$ & 0 & 1 & 0 & 1 & -- & 1 & 0 &M S$_0$\\
\hline
\end{tabular}
}
\label{table_9}
\end{center}
\end{table}

\begin{table}[H]
\caption{Type II superstrings: future asymptotic solutions
($t\rightarrow \infty$). M means a Milne type space.}
\begin{center}
{\footnotesize
\begin{tabular}{|c||c||c|c||c|c|c|c||c|c|c||c|}
\hline
&$\epsilon$&$\sigma_p$&$\sigma_q$&$\mu$&$\nu$&$a_0$&$b_0$&
$\lambda$&$\phi_1$&$t_E$& type \\
\hline
IIF1&0 & $1$ & $\pm 1$ & $0$ & 0.6201 & 1.078 & -- & 1
&0.2067&$\rightarrow \infty$&IIE6$_+$\\
\hline
IIF2&0 & $\pm 1$ & 1 & $0.7655$ & 0 & -- & 1.480& $e^{\mu t_E}$ & 0
&$\rightarrow \infty$&IIE4$_+$\\
\hline
IIF3&0 & $\pm 1$ & $-1$ & $0.6200$ & 0 & -- & 2.763 &$e^{\mu t_E}$ & 0
&$\rightarrow \infty$&IIE5$_+$\\
\hline
IIF4&0 & $\pm 1$ & $\pm 1$ & $0.7999$ & 0.1299 & -- & -- & 3.053 & 0.0433
&$\rightarrow \infty$&IIE1$_+$\\
\hline
IIF5&0 & $\pm 1$ & $\pm 1$ & $0.5075$ & 0.5075 & -- & --
&1.333&0.1692&$\rightarrow \infty$&IIE2$_+$
\\
\hline
IIF6&0 & $\pm 1$ & $\pm 1$ & $0.4962$ & 0.5131 & -- & -- &1.322 & 0.1710
&$\rightarrow \infty$&IIE3$_+$\\
\hline
\hline
IIF7&1&0&0&0.5556&$-0.1111$& -- & -- &0.3333&$-0.1667$&$\to \infty$&Kasner\\
\hline
IIF8&1&0&0&$-0.3333$&0.3333& -- &--&0.3333&0.1667&$\to \infty$&Kasner\\
\hline
IIF9&1&$-1$&$-1$&1&1&0.5&0.7906&1 &0.25 &$\rightarrow \infty$& M M\\
\hline
\end{tabular}
}
\label{table_10}
\end{center}
\end{table}

\begin{table}[H]
\caption{Type II superstrings: past asymptotic solutions
 ($t\rightarrow -\infty$).  K, S$_\pm$, S$_0$, M  and C mean
a kinetic dominance, a static space with positive (or negative) curvature,
a flat static space,  a Milne type space, and a constant curvature space,
respectively.}
\begin{center}
{\footnotesize
\begin{tabular}{|c||c||c|c||c|c|c|c||c|c|c||c|}
\hline
&$\epsilon$ & $\sigma_p$ & $\sigma_q$ & $\mu$ & $\nu$ & $a_0$ & $b_0$ &
$\lambda$ & $\phi_1$ & $t_E$ & type \\
\hline
IIP1 & 0 & $1$ & $\pm 1$ & 0 & $-0.6201$ & 1.078 & -- & 1 & $-0.2067$
&$\rightarrow -\infty$&IIE6$_-$\\
\hline
IIP2 & 0 & $\pm 1$ & $-1$ & $-0.6200$ & $0$ & -- & 2.763 &  $e^{\mu t_E}$ &
$0$ & $\rightarrow -\infty$ & IIE4$_-$ \\
\hline
IIP3 & 0 & $\pm 1$ & $1$ & $-0.7655$ & $0$ & -- & 1.480 & $e^{\mu t_E}$& $0$
&$\rightarrow -\infty$&IIE5$_-$\\
\hline
IIP4 & 0 & $\pm 1$ & $\pm 1$ & $-0.7999$ & $-0.1299$ & -- & -- & 3.053 &
$-0.0433 $ &$\rightarrow -\infty$&IIE1$_-$\\
\hline
IIP5 & 0 & $\pm 1$ & $\pm 1$ & $-0.5075$ & $-0.5075$ & -- & -- & 1.333 & $
-0.1692$ & $\rightarrow -\infty$ & IIE2$_-$ \\
\hline
IIP6 & 0 & $ \pm 1$ & $\pm 1$ & $-0.4962$ & $-0.5131$ & -- & -- & 1.322 & $
-0.1710 $ & $\rightarrow -\infty $ &IIE3$_-$ \\
\hline
\hline
IIP7&1 & $0,\pm 1$ & $0,\pm 1$ & $5.7427$ & $5.7427$ & -- &-- & 1.2602 &
0.3150 &$\sim 0$& K K\\
\hline
IIP8&1 & $0,\pm 1$ & $0,\pm 1$ & $0.3205$ & $0.00017$ & -- &-- & 0.3208 &
0.0001699 & $\sim 0$ & K K
\\
\hline
IIP9&1 & $0,\pm 1$ & $0,\pm 1$ & $0.2883$ & $0.2883$ & -- & -- & 0.6184 &
0.1546 & $\sim 0$ & K K
\\
\hline
IIP10&1 & $0,\pm 1$ & $0,\pm 1$ & $0.0013$ & $0.2954$ & -- & -- & 0.4705 &
0.1566 & $\sim 0$ & K K
\\
\hline
IIP11&1 & 0 & $1$ & 4.0305 & 1 & -- & 0.3375 & 1.7576 & 0.25 &$\sim 0$&  K C\\
\hline
IIP12&1 & 0 & $-1$ & 0.4484 & 1 & -- & 0.8948 & 0.8621 & 0.25  &$\sim 0$& K M\\
\hline
IIP13&1 & 0 & $-1$ & $-9.7439$ & 1 & -- & 0.1028 & $-1.6860$ & 0.25  &$\sim 0$
& K M\\
\hline
IIP14&1 & 1 & 0 & 1 & 6.1725 & 0.1148 & -- & 1 & 0.3163 &$\sim 0$& C K\\
\hline
IIP15&1 & $-1$ & 0 & 1 & 0.0358 & 0.9915 & -- & 1 & 0.0323 &$\sim 0$& M K\\
\hline
IIP16&1 & $-1$ & 0 & 1 & $-26.8744$ & 0.0323 & -- & 1 & 0.3375 &$\sim 0$& M K\\
\hline
IIP17&1 & $-1$ & $\pm 1$ & 1 & 0 & 1 & -- & 1 & 0  &$\sim 0$&M K \\
\hline
IIP18&1 & $\pm 1$ & $-1$ & 0 & 1 & -- & 1 & 0.75 & 0.25  &$\sim 0$&K M\\
\hline
IIP19&1 & 1 & 1 & 1 & 1 & 0.4698 & 0.6843 & 1 & 0.25  &$\sim 0$&C C \\
\hline
IIP20&1 & $-1$ & $-1$ & 1 & 1 & 1.127 & 0.7757 & 1 & 0.25  &$\sim 0$&M M\\
\hline
IIP21&1 & $-1$ & $-1$ & 1 & 1 & 0.6527 & 0.8733 & 1 &  0.25 &$\sim 0$&M M\\
\hline
IIP22&1 & $-1$ & 1 & 1 & 1 & 0.1701 & 0.2330 & 1 & 0.25  & $\sim 0$&M C\\
\hline
\end{tabular}
}
\label{table_11}
\end{center}
\end{table}

\newcommand{\NP}[1]{Nucl.\ Phys.\ B\ {\bf #1}}
\newcommand{\PL}[1]{Phys.\ Lett.\ B\ {\bf #1}}
\newcommand{\CQG}[1]{Class.\ Quant.\ Grav.\ {\bf #1}}
\newcommand{\CMP}[1]{Comm.\ Math.\ Phys.\ {\bf #1}}
\newcommand{\IJMP}[1]{Int.\ Jour.\ Mod.\ Phys.\ {\bf #1}}
\newcommand{\JHEP}[1]{JHEP\ {\bf #1}}
\newcommand{\PR}[1]{Phys.\ Rev.\ D\ {\bf #1}}
\newcommand{\PRL}[1]{Phys.\ Rev.\ Lett.\ {\bf #1}}
\newcommand{\PRE}[1]{Phys.\ Rep.\ {\bf #1}}
\newcommand{\PTP}[1]{Prog.\ Theor.\ Phys.\ {\bf #1}}
\newcommand{\PTPS}[1]{Prog.\ Theor.\ Phys.\ Suppl.\ {\bf #1}}
\newcommand{\MPL}[1]{Mod.\ Phys.\ Lett.\ {\bf #1}}
\newcommand{\JP}[1]{Jour.\ Phys.\ {\bf #1}}

\end{document}